\documentclass{article}

\usepackage[table]{xcolor}

\usepackage[final]{neurips_data_2022}
\usepackage{subfigure}

\DeclareRobustCommand\onedot{\futurelet\@let@token\@onedot}
 
\def\ie{\emph{i.e.}}

\usepackage[utf8]{inputenc} 
\usepackage[T1]{fontenc}    
\usepackage{url}            
\usepackage{booktabs}       
\usepackage{amsfonts}       
\usepackage{nicefrac}       
\usepackage{microtype}      
\usepackage{xcolor}         
\usepackage{graphicx}
\usepackage{enumitem}
\usepackage{color}
\usepackage{multirow}
\usepackage{makecell}
\usepackage{hyperref}

\title{TwiBot-22: Towards Graph-Based \\ Twitter Bot Detection}

\author{
Shangbin Feng$^{1,2}$\thanks{\ \ These authors contributed equally to this work.} \ \ Zhaoxuan Tan$^1$\footnotemark[1] \ \ Herun Wan$^1$\footnotemark[1] \ \ Ningnan Wang$^1$\footnotemark[1] \ \ Zilong Chen$^{1,3}$\footnotemark[1] \ \ Binchi Zhang$^{1,4}$\footnotemark[1] \vspace{3pt} \\ 
\textbf{Qinghua Zheng$^1$\thanks{\ \ Corresponding author: Qinghua Zheng, School of Computer Science and Technology, Xi'an Jiaotong University. Email: qhzheng@xjtu.edu.cn; Minnan luo, School of Computer Science and Technology, Xi'an Jiaotong University. Email: minnluo@xjtu.edu.cn} \ \ Wenqian Zhang$^1$ \ \ Zhenyu Lei$^1$ \ \ Shujie Yang$^1$ \ \ Xinshun Feng$^1$ \ \ Qingyue Zhang$^1$} \vspace{3pt} \\ 
\textbf{Hongrui Wang$^1$ \ \ Yuhan Liu$^1$ \ \ Yuyang Bai$^1$ \ \ Heng Wang$^1$ \ \ Zijian Cai$^1$ \ \ Yanbo Wang$^1$} \vspace{3pt} \\ 
\textbf{Lijing Zheng$^1$ \ \ Zihan Ma$^1$ \ \ Jundong Li$^4$ \ \ Minnan Luo$^1$} \vspace{3pt} \\
Xi'an Jiaotong University$^1$, University of Washington$^2$, Tsinghua University$^3$, University of Virginia$^4$ \\
contact: \href{mailto:shangbin@cs.washington.edu}{\texttt{shangbin@cs.washington.edu}}
}

\begin{document}

\maketitle

\begin{abstract}
Twitter bot detection has become an increasingly important task to combat misinformation, facilitate social media moderation, and preserve the integrity of the online discourse. State-of-the-art bot detection methods generally leverage the graph structure of the Twitter network, and they exhibit promising performance when confronting novel Twitter bots that traditional methods fail to detect. However, very few of the existing Twitter bot detection datasets are graph-based, and even these few graph-based datasets suffer from limited dataset scale, incomplete graph structure, as well as low annotation quality. In fact, the lack of a large-scale graph-based Twitter bot detection benchmark that addresses these issues has seriously hindered the development and evaluation of novel graph-based bot detection approaches. In this paper, we propose TwiBot-22, a comprehensive graph-based Twitter bot detection benchmark that presents the largest dataset to date, provides diversified entities and relations on the Twitter network, and has considerably better annotation quality than existing datasets. In addition, we re-implement 35 representative Twitter bot detection baselines and evaluate them on 9 datasets, including TwiBot-22, to promote a fair comparison of model performance and a holistic understanding of research progress. To facilitate further research, we consolidate all implemented codes and datasets into the TwiBot-22 evaluation framework, where researchers could consistently evaluate new models and datasets. The TwiBot-22 Twitter bot detection benchmark and evaluation framework are publicly available at \url{https://twibot22.github.io/}.
\\
\end{abstract}

\section{Introduction}
Automated users on Twitter, also known as Twitter bots, have become a widely known and well-documented phenomenon. Over the past decade, malicious Twitter bots were responsible for a wide range of problems such as online disinformation \citep{cui2020deterrentmisinfo1, wang2020fakemisinfo2, lu2020gcanmisinfo3}, election interference \citep{howard2016botselection1, neudert2017junkelection2, rossi2020detectingelection3, ferrara2017disinformationelection4}, extremism campaign \citep{ferrara2016predictingextremism1, marcellino2020counter}, and even the spread of conspiracy theories \citep{ferrara2020typesconspiracy1, ahmed2020covidconspiracy2, anwar2021analyzingconspiracy3}. These societal challenges have called for automatic Twitter bot detection models to mitigate their negative influence.

Existing Twitter bot detection models are generally \textbf{feature-based}, where researchers propose to extract numerical features from user information such as metadata \citep{yang2020scalableyangetal, lee2011sevenleeetal}, user timeline \citep{mazza2019cresci-rtbust-2019, chavoshi2016debot}, and follow relationships \citep{beskow2020youbeskowetalfriend, chu2012detectingreputation}. However, feature-based approaches are susceptible to adversarial manipulation, \ie, when bot operators try to avoid detection by tampering with these hand-crafted features \citep{cresci-2017aparadigm, cresci2020decade}. Researchers also proposed \textbf{text-based} approaches, where text analysis techniques such as word embeddings \citep{wei2019twitterweietal}, recurrent neural networks \citep{kudugunta2018deepkuduguntaetal, feng2021satar}, and pre-trained language models \citep{dukic2020youdukicetal} are leveraged to analyze tweet content and identify malicious intent. However, new generations of Twitter bots often intersperse malicious content with normal tweets stolen from genuine users \citep{cresci2020decade, feng2021botrgcn}, thus their bot nature becomes more subtle to text-based methods. With the advent of graph neural networks, recent advances focus on developing \textbf{graph-based} Twitter bot detection models. These methods \citep{Alhosseinietal,feng2021botrgcn} interpret users as nodes and follow relationships as edges to leverage graph mining techniques such as GCN \citep{kipf2016semigcn}, R-GCN \citep{schlichtkrull2018modelingrgcn}, and RGT \citep{RGT} for graph-based bot detection. In fact, recent research have shown that graph-based approaches achieve state-of-the-art performance, are capable of detecting novel Twitter bots, and are better at addressing various challenges facing Twitter bot detection \citep{feng2021botrgcn, RGT}.


\begin{figure}[t]
    \centering
    \includegraphics[width=1\textwidth]{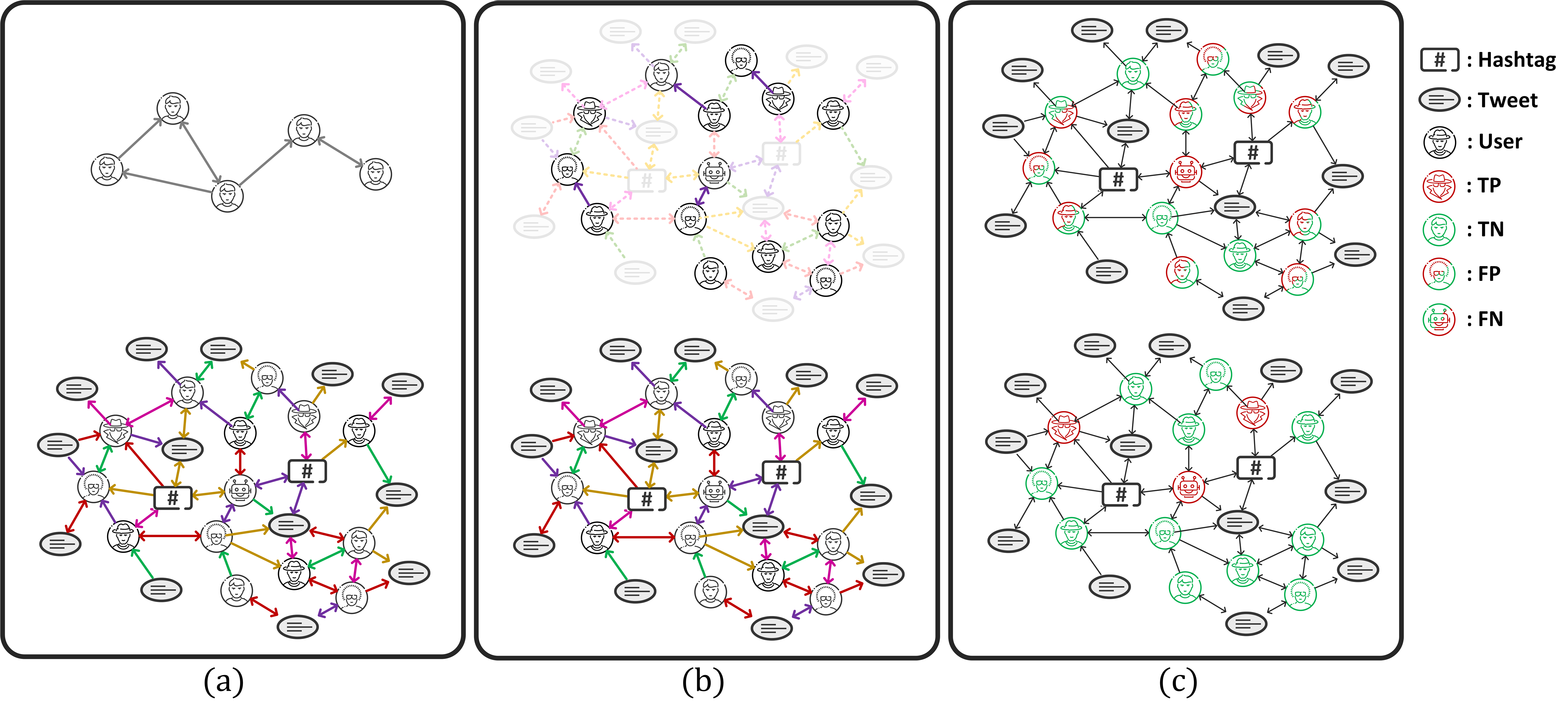}
    \caption{Compared to real-world Twitter (below), existing Twitter bot detection datasets (above) suffer from (a) limited dataset scale, (b) incomplete graph structure, and (c) poor annotation quality.}
    \label{fig:teaser}
\end{figure}

However, the development and evaluation of graph-based Twitter bot detection models are poorly supported by existing datasets. The Bot Repository\footnote{\url{https://botometer.osome.iu.edu/bot-repository/datasets.html}} provides a comprehensive collection of existing datasets. Out of the 18 listed datasets, only two of them, TwiBot-20 \citep{feng2021twibot-20} and cresci-2015 \citep{cresci-2015fame}, explicitly provide the graph structure among Twitter users. In addition, these two graph-based datasets suffer from the following issues as illustrated in Figure \ref{fig:teaser}:
\begin{itemize}[leftmargin=*]
    \item \textbf{(a) limited dataset scale}. Twibot-20 \citep{feng2021twibot-20} contains 11,826 labeled users and cresci-15 \citep{cresci-2015fame} contains 7,251 labeled users, while online conversations and discussions about heated topics often involve hundreds of thousands of users \citep{banda2021large}.
    \item \textbf{(b) incomplete graph structure}. Real-world Twitter is a heterogeneous information network that contains many types of entities and relations \citep{RGT}, while TwiBot-20 and cresci-15 only provide users and follow relationships.
    \item \textbf{(c) low annotation quality}. TwiBot-20 resorted to crowdsourcing for user annotation, while crowdsourcing leads to significant noise \citep{graells2022bots20labelquality} and is susceptible to the false positive problem \citep{rauchfleisch2020false}.
\end{itemize}

In light of these challenges, we propose TwiBot-22, a graph-based Twitter bot detection benchmark that addresses these issues. Specifically, TwiBot-22 adopts a two-stage controlled expansion to sample the Twitter network, which results in a dataset that is 5 times the size of the largest existing dataset. TwiBot-22 provides 4 types of entities and 14 types of relations in the Twitter network, which provides the first (truly) heterogeneous graph for Twitter bot detection. Finally, TwiBot-22 adopts the weak supervision learning strategy for data annotation which results in significantly improved annotation quality. To compare TwiBot-22 with existing datasets, we re-implement 35 Twitter bot detection baselines and evaluate them on 9 datasets, including TwiBot-22, to provide a holistic view of research progress and highlight the advantages of TwiBot-22. We consolidate all datasets and implemented codes into the TwiBot-22 evaluation framework to facilitate further research. Our main contributions are summarized as follows:
\begin{itemize}[leftmargin=*]
    \item We propose TwiBot-22, a graph-based Twitter bot detection dataset that establishes the largest benchmark to date, provides diversified entities and relations in the Twitter network, and has considerably improved annotation quality.
    \item We re-implement and benchmark 35 existing Twitter bot detection models on 9 datasets, including TwiBot-22, to compare different approaches fairly and facilitate a holistic understanding of research progress in Twitter bot detection.
    \item We present the TwiBot-22 evaluation framework, where researchers could easily reproduce our results, examine existing datasets and methods, infer on unseen Twitter data, and contribute new datasets and models to the framework.
\end{itemize}

\section{Related Work}

\subsection{Twitter Bot Detection}
Existing Twitter bot detection methods mainly fall into three categories: \textbf{feature-based} methods, \textbf{text-based} methods, and \textbf{graph-based} methods.

\textbf{Feature-based} methods conduct feature engineering with user information and apply traditional classification algorithms for bot detection. Various features are extracted from user metadata \citep{kudugunta2018deepkuduguntaetal}, tweets \citep{miller2014twittermilleretal}, description \citep{hayawi2022deeprobothayawietal}, temporal patterns \citep{mazza2019cresci-rtbust-2019}, and follow relationships \citep{feng2021satar}. Later research efforts improve the scalability of feature-based approaches \citep{yang2020scalableyangetal}, automatically discover new bots \citep{chavoshi2016debot}, and strike the balance between precision and recall \citep{morstatter2016newmorstatteretal}. However, as bot operators are increasingly aware of these hand-crafted features, novel bots often try to tamper with these features to evade detection \citep{cresci2020decade}. As a result, feature-based methods struggle to keep up with the arms race of bot evolution \citep{feng2021satar}.

\textbf{Text-based} methods use techniques in natural language processing to detect Twitter bots based on tweets and user descriptions. Word embeddings \citep{wei2019twitterweietal}, recurrent neural networks \citep{kudugunta2018deepkuduguntaetal}, the attention mechanism \citep{feng2021satar}, and pre-trained language models \citep{dukic2020youdukicetal} are adopted to encode tweets for bot detection. Later research combines tweet representations with user features \citep{cai2017detectingcaietal}, learns unsupervised user representations \citep{feng2021satar}, and attempts to address the multi-lingual issue in tweet content \citep{knauth2019languageknauthetal}. However, novel bots begin to counter text-based approaches by diluting malicious tweets with content stolen from genuine users \citep{cresci2020decade}. In addition, \citet{feng2021botrgcn} shows that analyzing tweet content alone might not be robust and accurate for bot detection.

\textbf{Graph-based} methods interpret Twitter as graphs and leverage concepts from network science and geometric deep learning for Twitter bot detection. Node centrality \citep{dehghan2022detectingdehghanetal}, node representation learning \citep{pham2022bot2vec}, graph neural networks (GNNs) \citep{Alhosseinietal}, and heterogeneous GNNs \citep{feng2021botrgcn} are adopted to conduct graph-based Twitter bot detection. Later research try to combine graph-based and text-based methods \citep{guo2021socialbgsrd} or propose new GNN architectures to leverage heterogeneities in the Twitter network \citep{RGT}. Graph-based approaches have shown great promise in tackling various challenges facing Twitter bot detection, such as bot communities and bot disguise \citep{feng2021botrgcn}.

The development and evaluation of these models would not be possible without the many valuable Twitter bot detection datasets that were proposed over the past decade. These datasets mainly focus on politics and elections in the United States \citep{yang2020scalableyangetal} and European countries \citep{cresci-2017aparadigm}. Cresci-17 \citep{cresci-2017aparadigm} propose the concept of "social spambots" and presents a widely used dataset with more than one type of bots. TwiBot-20 \citep{feng2021twibot-20} is the latest and most comprehensive Twitter bot detection dataset that addresses the issue of user diversity in previous datasets. However, among 18 datasets presented in the Bot Repository, the go-to place for Twitter bot detection research, only 2 explicitly provide the graph structure of the Twitter network. In addition, these datasets suffer from limited dataset scale, incomplete graph structure, and low annotation quality while increasingly falling short of consistently benchmarking novel graph-based approaches. In light of these challenges, we present TwiBot-22 to alleviate these issues, promote a rethinking of research progress, and facilitate further research in Twitter bot detection.

\subsection{Graph-based Social Network Analysis}
Users in online social networks interact with each other and become part of the network structure, while the network structure is essential in understanding the patterns of social media \citep{carrington2005models}. With the advent of geometric deep learning, graph neural networks (GNNs) have become increasingly popular in social network analysis research. \citet{qian2021distilling} propose to model social media with heterogeneous graphs and leverage relational GNNs for illicit drug trafficker detection. \citet{guo2021dual} propose dual graph enhanced embedding neural network to improve graph representation learning and tackle challenges in click-through rate prediction. Graphs and GNNs are also adopted to detect online fraud \citep{liu2021intentionfraud1, li2021livefraud2, wang2021modelingfraud3,mishra-2021-modeling-users, dou2020enhancingfraud4}, combat misinformation \citep{cui2020deterrentmisinfo1, wang2020fakemisinfo2, lu2020gcanmisinfo3, varlamis2022survey, hu2021compare}, and improve recommender systems \citep{ying2018graphrecommend1, wu2020graphrecommend2}. The task of Twitter bot detection is no exception, where novel and state-of-the-art approaches are increasingly graph-based \citep{Alhosseinietal, GraphHist, feng2021botrgcn, RGT, lei2022bic}. In this paper, we propose the TwiBot-22 benchmark to better support the development and evaluation of graph-based Twitter bot detection models.

\section{TwiBot-22 Dataset}

\subsection{Data Collection}
TwiBot-22 aims to present a large-scale and graph-based Twitter bot detection benchmark. To this end, we adopt a two-stage data collection process. We firstly adopt diversity-aware breadth-first search (BFS) to obtain the user network of TwiBot-22. We then collect additional entities and relations on the Twitter network to enrich the heterogeneity of the TwiBot-22 network.

\paragraph{User network collection.} A common drawback of existing Twitter bot detection datasets is that they only feature a few types of bots and genuine users, while real-world Twitter is home to diversified users and bots \citep{feng2021twibot-20}. As a result, TwiBot-20 proposes to use breadth-first search (BFS) for user collection, starting from "seed users" and expanding with user follow relationships. To ensure that TwiBot-22 includes different types of bots and genuine users, we augment BFS with two diversity-aware sampling strategies:

\begin{itemize}[leftmargin=*]
    \item \textbf{Distribution diversity}. Given user metadata such as follower count, different types of users fall differently into the metadata distribution. Distribution diversity aims to sample users in the top, middle, and bottom of the distribution. For numerical metadata, among a user's neighbors and their metadata values, we select the $k$ users with the highest value, $k$ with the lowest, and $k$ randomly chosen from the rest. For true-or-false metadata, we select $k$ with true and $k$ with false.
    \item \textbf{Value diversity}. Given a user and its metadata, value diversity is adopted so that neighbors with significantly different metadata values are more likely to be included, ensuring the diversity of collected users. For numerical metadata, among expanding user $u$'s neighbors $X$ and their metadata values $x^{num}$, the probability of user $x \in X$ being sampled is denoted as $p(x) \propto |u^{num} - x^{num}|$. For true-or-false metadata we select $k$ users from the opposite class.
\end{itemize}

Based on these sampling strategies, TwiBot-22 conducts diversity-aware BFS starting from \textit{@NeurIPSConf}. For each neighborhood expansion, one metadata and one of the sampling strategies are randomly adopted. The BFS process stops until the user network contains a desirable amount of Twitter users. More information about the user collection process is presented in Appendix \ref{app:collection_detail}.

\paragraph{Heterogeneous graph building.} The user network collection process constructs a homogeneous graph with users as nodes and follow relationships as edges. Apart from that, the Twitter network contains diversified entities and relations such as lists and retweets. Based on the user network, we collect the tweets, associated lists, and mentioned hashtags of these users as well as 12 other relations between users and these new entities. As a result, TwiBot-22 presents a heterogeneous Twitter network with 4 types of entities and 14 types of relations. More information about the heterogeneous Twitter network is presented in Appendix \ref{app:dataset_stat}.

As a result, we obtain the TwiBot-22 heterogeneous graph that contains 92,932,326 nodes and 170,185,937 edges. We present more dataset statistics in Table \ref{tab:statistic} in the appendix.



\subsection{Data Annotation}Existing bot detection datasets often rely on manual annotation or crowdsourcing, while it is labor-intensive and thus not feasible with the large-scale TwiBot-22. As a result, we adopt weak supervision learning strategy to generate high-quality labels. We firstly invite bot detection experts to annotate 1,000 Twitter users in TwiBot-22. We then generate noisy labels with the help of bot detection models. Finally, we generate high-quality annotations for TwiBot-22 with the Snorkel framework \citep{ratner2017snorkel}.

\paragraph{Expert annotation.} We randomly select 1,000 users in TwiBot-22 and assign each user to 5 Twitter bot detection experts to identify if it is a bot. We then create an 8:2 split for these expert annotations as training and test sets. More details about these experts are presented in Appendix \ref{app:expert_annotation}.
\label{appendix: test1}
\begin{figure}[t]
    \centering
    \includegraphics[width=1\linewidth]{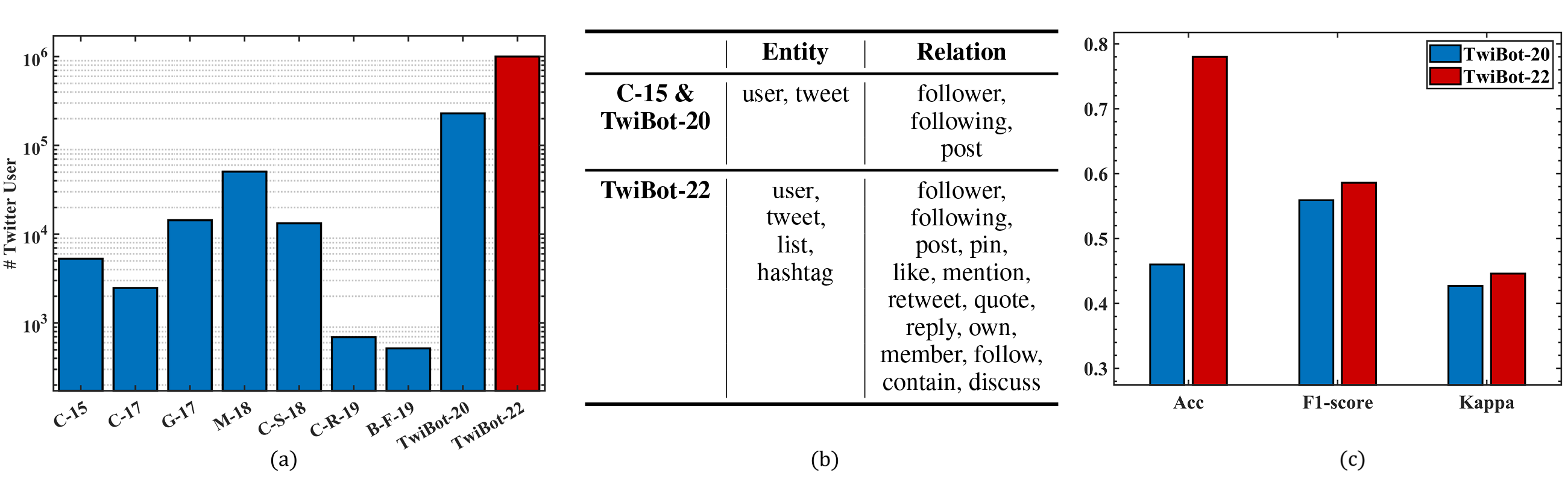}
    \caption{Analyzing TwiBot-22. (a) Dataset scale in terms of users. (b) List of entities and relations in TwiBot-22 and graph-based datasets. (c) Accuracy and F1-score of dataset labels compared to expert annotations as well as Randolph's kappa coefficient \citep{randolph2005free} of expert agreement.}
    \label{fig:data_analysis}
\end{figure}

\paragraph{Generate noisy labels.} We employ 8 hand-crafted labeling functions and 7 competitive feature-based and neural network-based bot detection models to generate noisy labels. For handcrafted labeling functions, we adopt spam keywords in tweets and user descriptions as well as user categorical features such as verified and sensitive tweets. For feature engineering models, we select features based on users' metadata such as creation time, follower count and name length. We then adopt Adaboost, random forest, and MLP to result in three feature-based classifiers. For neural network-based models, we follow \citet{feng2021botrgcn} to encode user information and employ MLP, GAT \citep{velivckovic2017graphgat}, GCN \citep{kipf2016semigcn}, and R-GCN \citep{schlichtkrull2018modelingrgcn} as four classifiers. We train these classifiers on the training set of expert annotations and calculate the uncertainty scores for all users in TwiBot-22 under each classifier as $\phi=-(\hat{y}_0\log(\hat{y}_0)+\hat{y}_1\log(\hat{y}_1))$, where $\hat{y}_0$ and $\hat{y}_1$ denote the probability of being genuine users or bots. For each classifier, we then remove model predictions with the top 40\% uncertainty scores to alleviate label noises.

\paragraph{Majority voting.} After obtaining the noisy labels, we evaluate their plausibility with Snorkel \citep{ratner2017snorkel} and clean the labels at the same time. The output of the Snorkel system are probabilistic labels, thus we use these labels to train an MLP classifier to obtain the final annotations of TwiBot-22. We further evaluate the annotation quality on the test set of expert annotations and we obtain an 90.5\% accuracy. Compared to the 80\% accuracy standard in TwiBot-20 \citep{feng2021twibot-20}, TwiBot-22 has considerably imporved annotation quality.

\subsection{Data Analysis}
Existing graph-based Twitter bot detection datasets suffer from limited dataset scale, incomplete graph structure, and low annotation quality. As a result, we examine whether TwiBot-22 has adequately addressed these challenges and present our findings in Figure \ref{fig:data_analysis}.

         

\paragraph{Dataset scale.} Figure \ref{fig:data_analysis}(a) illustrates the number of Twitter users in TwiBot-22 and existing datasets. It is illustrated that TwiBot-22 establishes the largest Twitter bot detection benchmark to date, with approximately 5 times more users than the second-largest TwiBot-20.

\paragraph{Graph structure.} Figure \ref{fig:data_analysis}(b) demonstrates that the TwiBot-22 network contains 4 types of entities and 14 types of relations, resulting in significantly enriched graph structure compared to existing graph-based datasets cresci-15 and TwiBot-20 with only 2 entity types and 3 relation types.


\paragraph{Annotation quality.} TwiBot-20, the largest graph-based Twitter bot detection benchmark to date, leveraged crowdsourcing for data annotation. To improve label quality, TwiBot-22 uses weak supervision learning strategies and leverages 1,000 expert annotations to guide the process. To examine whether TwiBot-22 has improved annotation quality than TwiBot-20, we ask Twitter bot detection experts to participate in an "expert study", where they are asked to evaluate users in TwiBot-20 and TwiBot-22 to examine how often do experts agree with dataset labels. Figure \ref{fig:data_analysis}(c) illustrates the results, which shows that these experts find TwiBot-22 to provide more consistent, accurate, and trustworthy data annotations. More details about the expert study are available in Appendix \ref{app:annotation_quality}.
\label{appendix: test2}
\section{Experiments}

\subsection{Experiment Settings}

\paragraph{Datasets.} We evaluate Twitter bot detection models on all 9 datasets in the Bot Reporistory that contain both bots and genuine users: cresci-2015 \citep{cresci-2015fame}, gilani-2017 \citep{gilani-2017bots}, cresci-2017 \citep{cresci-2017aparadigm, cresci-2017bsocial}, midterm-18 \citep{yang2020scalableyangetal}, cresci-stock-2018 \citep{cresci-stock-2018afake, cresci-stock-2018bcashtag}, cresci-rtbust-2019 \citep{mazza2019cresci-rtbust-2019}, botometer-feedback-2019 \citep{yang2019armingbotometer-feedback-2019}, TwiBot-20 \citep{feng2021twibot-20}, and TwiBot-22. We present dataset details in Table \ref{tab:dataset_details}. We create a 7:2:1 random split as training, validation, and test set to ensure fair comparison.

\begin{table}[t]
    \caption{Statistics of the 9 datasets. TwiBot-20 contains unlabelled users so that \# User $\neq$ \# Human $+$ \# Bot. C-15, G-17, C-17, M-18, C-S-18, C-R-19, B-F-19 are short for cresci-2015, gilani-2017, cresci-2017, midterm-18, cresci-stock-2018, cresci-rtbust-2019, botometer-feedback-2019. C-17 contains only "post" edges between users and tweets, which is not a graph-based dataset.}
    \label{tab:dataset_details}
\renewcommand\arraystretch{1.2}
    \centering
    \resizebox{\linewidth}{!}{
    \begin{tabular}{c|c|c|c|c|c|c|c|c|c} 
         \toprule[1.5pt]
         \textbf{Dataset} & \textbf{C-15} &\textbf{G-17} & \textbf{C-17} & \textbf{M-18} & \textbf{C-S-18} & \textbf{C-R-19} & \textbf{B-F-19} & \textbf{TwiBot-20} & \textbf{TwiBot-22}  \\ \midrule[1pt]
         \textbf{\# Human} & 1,950 & 1,394 & 3,474 & 8,092 & 6,174 & 340 & 380 & 5,237 & 860,057 \\
         \textbf{\# Bot} & 3,351 & 1,090 & 10,894 & 42,446 & 7,102 & 353 & 138 & 6,589 & 139,943 \\
         \textbf{\# User} & 5,301 & 2,484 & 14,368 & 50,538 & 13,276 & 693 & 518 & 229,580 & 1,000,000  \\ 
         \textbf{\# Tweet} & 2,827,757 & 0 & 6,637,615 & 0 & 0 & 0 & 0 & 33,488,192 & 88,217,457 \\
         \textbf{\# Human Tweet} & 2,631,730 & 0 & 2,839,361 & 0 & 0 & 0 & 0 & 33,488,192 & 81,250,102 \\
         \textbf{\# Bot Tweet} & 196,027 & 0 & 3,798,254 & 0 & 0 & 0 & 0 & 33,488,192 & 6,967,355 \\
         \textbf{\# Edge} & 7,086,134 & 0 & 6,637,615 & 0 & 0 & 0 & 0 & 33,716,171 & 170,185,937 \\
         \bottomrule[1.5pt]
    \end{tabular}
    }
\end{table}

\paragraph{Baselines.} We re-implement and evaluate 35 Twitter bot detection baselines SGBot \citep{yang2020scalableyangetal}, \citet{kudugunta2018deepkuduguntaetal}, \citet{hayawi2022deeprobothayawietal}, BotHunter \citep{beskow2018botbothunter}, NameBot \citep{beskow2019itsbeskowetalname}, \citet{abreu2020twitterabreuetal}, \citet{cresci2016dna1}, \citet{wei2019twitterweietal}, BGSRD \citep{guo2021socialbgsrd}, RoBERTa \citep{liu2019roberta}, T5 \citep{raffel2019exploringt5}, \citet{efthimion2018supervisedefthimionetal}, \citet{kantepe2017preprocessingkantepeetal}, \citet{miller2014twittermilleretal}, \citet{varol2017onlinevaroletal}, \citet{kouvela2020botbotdetective}, \citet{ferreira2019uncoveringsantosetal}, \citet{lee2011sevenleeetal}, LOBO \citep{echeverri2018lobo}, \citet{moghaddam2022friendshipmoghaddametal}, \citet{Alhosseinietal}, \citet{knauth2019languageknauthetal}, FriendBot \citep{beskow2020youbeskowetalfriend}, SATAR \citep{feng2021satar}, Botometer \citep{yang2022botometer}, \citet{rodriguez2020onerodriguezetal}, GraphHist \citep{GraphHist}, EvolveBot \citep{yang2013empiricalevolvebot}, \citet{dehghan2022detectingdehghanetal}, GCN \citep{kipf2016semigcn}, GAT \citep{velivckovic2017graphgat}, HGT \citep{hu2020heterogeneoushgt}, SimpleHGN \citep{lv2021wesimplehgn}, BotRGCN \citep{feng2021botrgcn}, RGT \citep{RGT}. These methods leverage different aspects of user information and represent different stages of bot detection research. More details about these baseline methods are available in Appendix \ref{app:baseline_detail}.

\begin{table*}[t]
    \caption{Average bot detection accuracy and standard deviation of five runs of 35 baseline methods on 9 datasets. \textbf{Bold} and \underline{underline} indicate the highest and second highest performance. The F, T, and G in the "Type" column indicates whether a baseline is feature-based, text-based, or graph-based. Cresci \textit{et al.} and Botometer are deterministic methods or APIs without standard deviation. "/" indicates that the dataset does not contain enough user information to support the baseline. "-" indicates that the baseline is not scalable to the largest TwiBot-22 dataset.}
    \label{tab:big_acc}
\renewcommand\arraystretch{1.3}
    \centering
    \resizebox{1\linewidth}{!}{
    \begin{tabular}{ccccccccccc}
        \toprule[1.5pt]
        \textbf{Method} & \textbf{Type} & \textbf{C-15} & \textbf{G-17} & \textbf{C-17} & \textbf{M-18} & \textbf{C-S-18} & \textbf{C-R-19} & \textbf{B-F-19} & \textbf{TwiBot-20} & \textbf{TwiBot-22} \\ 
        \midrule[1pt]
        
        SGBot & F & $77.1 ~(\pm0.2)$ & $\textbf{78.6} ~(\pm0.8)$ & $92.1 ~(\pm0.3)$ & $\underline{99.2} ~(\pm0.0)$ & $81.3 ~(\pm0.1)$ & $80.9 ~(\pm1.5)$ & $75.5 ~(\pm1.9)$ & $81.6 ~(\pm0.5)$ & $75.1 ~(\pm0.1)$ \\%
        
        Kudugunta \textit{et al.} & F &$75.3 ~(\pm 0.1)$ & $70.0 ~(\pm 1.1)$ & $88.3 ~(\pm 0.2)$ & $91.0 ~(\pm 0.5)$ & $77.5 ~(\pm 0.1)$ & $62.9 ~(\pm 0.8)$ & $74.0 ~(\pm 4.7)$ & $59.6 ~(\pm 0.7)$ & $65.9 ~(\pm 0.0)$ \\%
        
       Hayawi \textit{et al.} & F & $84.3 ~(\pm0.0)$ & $52.7 ~(\pm0.0)$  & $90.8 ~(\pm0.0)$ & $84.6 ~(\pm0.0)$ & $50.0 ~(\pm0.0)$ & $51.2 ~(\pm0.0)$ & $\underline{77.0} ~(\pm0.0)$ & $73.1 ~(\pm0.0)$ & $76.5 ~(\pm0.0)$ \\%
        
        BotHunter & F & $96.5 ~(\pm1.2)$ & $\underline{76.4} ~(\pm1.0)$ & $88.1 ~(\pm0.2)$ & $\textbf{99.3} ~(\pm0.0)$ & $81.2 ~(\pm0.2)$ & $\underline{81.5} ~(\pm1.7)$ & $74.7 ~(\pm1.0)$ & $75.2 ~(\pm0.4)$ & $72.8 ~(\pm0.0)$ \\
        
        NameBot & F & $77.0 ~(\pm0.0)$ & $60.8 ~(\pm0.0)$  & $76.8 ~(\pm0.0)$ & $85.1 ~(\pm0.0)$ & $55.8 ~(\pm0.0)$ & $63.2 ~(\pm0.0)$ & $71.7 ~(\pm0.0)$ & $59.1 ~(\pm0.1)$ & $70.6~(\pm0.0)$\\
        
        Abreu \textit{et al.} & F & $75.7 ~(\pm0.1)$ & $74.3 ~(\pm0.1)$ & $92.7~(\pm0.1)$ & $96.5 ~(\pm0.1)$ & $75.4 ~(\pm0.1)$ & $80.9 ~(\pm0.1)$ & $\textbf{77.4} ~(\pm0.1)$ & $73.4~(\pm0.1)$ & $70.7 ~(\pm0.1)$ \\
        \midrule[1pt]
        
        Cresci \textit{et al.} & T & $37.0$ & / & $33.5$ & / & / & / & / & $47.8$ & - \\
        
        Wei \textit{et al.} & T & $96.1 ~(\pm1.4)$& /& $89.3 ~(\pm0.7)$ & /& /& / &/ & $71.3 ~(\pm1.6)$ & $70.2 ~(\pm1.2)$ \\%
        
        BGSRD & T & $87.8 ~(\pm0.6)$ & $48.5 ~(\pm8.4)$ & $75.9 ~(\pm0.0)$ & $82.9 ~(\pm1.5)$ & $50.7 ~(\pm1.3)$ & $50.0 ~(\pm4.9)$ & $59.6 ~(\pm3.1)$ & $66.4 ~(\pm1.0)$ & $71.9 ~(\pm1.8)$\\
        
        RoBERTa & T & $97.0 ~(\pm 0.1) $ & / & $97.2 ~(\pm 0.0)$ & / & / & / & / & $75.5 ~(\pm 0.1)$ & $72.1 ~(\pm 0.1)$ \\
        
        
        T5 & T & $92.3 ~(\pm 0.1)$ & / & $96.4 ~(\pm 0.0)$ & / & / & / & / & $73.5 ~(\pm 0.1)$ & $72.1 ~(\pm 0.1)$\\%
        \midrule[1pt]

        Efthimion \textit{et al.} & FT & $92.5 ~(\pm 0.0)$ & $55.5 ~(\pm 0.0)$ & $88.0 ~(\pm 0.0)$ & $93.4 ~(\pm 0.0)$ & $70.8 ~(\pm 0.0)$ & $67.6~(\pm 0.0)$ & $69.8~(\pm 0.0)$ & $62.8 ~(\pm 0.0)$ & $74.1 ~(\pm 0.0)$ \\%
        
        Kantepe \textit{et al.} & FT & $97.5 ~(\pm1.3)$ & / & $98.2 ~(\pm1.5)$ & / & / & /&/ & $80.3 ~(\pm4.3)$ & $76.4 ~(\pm2.4)$ \\%
        
        Miller \textit{et al.} & FT & $75.5 ~(\pm0.0)$ & $51.0 ~(\pm0.0)$ & $77.1 ~(\pm0.2)$ & $83.7 ~(\pm0.0)$ & $52.5 ~(\pm0.0)$ & $54.4 ~(\pm0.0)$ & $\textbf{77.4} ~(\pm0.0)$ & $64.5 ~(\pm0.4)$ & $30.4 ~(\pm0.1)$ \\

        Varol \textit{et al.} & FT & $93.2 ~(\pm0.5)$ & / & / & / & / & / & / & $78.7 ~(\pm0.6)$ & $73.9 ~(\pm0.0)$ \\
        
        Kouvela \textit{et al.} & FT & $97.8 ~(\pm0.5)$ & $74.7 ~(\pm0.9)$ & $\underline{98.4} ~(\pm0.1)$ & $97.0 ~(\pm0.1)$ & $79.3 ~(\pm0.3)$ & $79.7 ~(\pm1.2)$ & $71.3 ~(\pm0.9)$ & $84.0 ~(\pm0.4)$ & $76.4 ~(\pm0.0)$ \\%
        
        Santos \textit{et al.} & FT & $70.8 ~(\pm 0.0)$ & $51.4 ~(\pm 0.0)$ & $73.8 ~(\pm 0.0)$ & $86.6 ~(\pm 0.0)$ & $62.5 ~(\pm 0.0)$ & $73.5 ~(\pm 0.0)$ & $71.7 ~(\pm 0.0)$ & $58.7 ~(\pm 0.0)$ & - \\
        
        Lee \textit{et al.} & FT & $\underline{98.2} ~(\pm0.1)$ & $74.8 ~(\pm1.2)$ & $\textbf{98.8} ~(\pm0.1)$ & $96.4 ~(\pm0.1)$ & $\underline{81.5} ~(\pm0.4)$ & $\textbf{83.5} ~(\pm1.9)$ & $75.5 ~(\pm1.3)$ & $77.4 ~(\pm0.5)$ & $76.3 ~(\pm0.1)$ \\%
        
        LOBO & FT & $\textbf{98.4} ~(\pm 0.3)$ & / & $96.6 ~(\pm 0.3)$ & / & / & / & / & $77.4~(\pm 0.2)$ & $75.7 ~(\pm 0.1)$ \\
        \midrule[1pt]

        Moghaddam \textit{et al.} & FG & $73.6 ~(\pm0.2)$ & / & / & / & / & / & / & $74.0 ~(\pm0.8)$ & $73.8 ~(\pm0.0)$ \\
        
        Alhosseini \textit{et al.} & FG & $89.6 ~(\pm 0.6)$ & / & / & / & / & / & / & $59.9 ~(\pm 0.6)$ & $47.7 ~(\pm 8.7)$ \\
        \midrule[1pt]
        
        Knauth \textit{et al.} & FTG & $85.9 ~(\pm0.0)$ & $49.6 ~(\pm0.0)$ & $90.2 ~(\pm0.0)$ & $83.9 ~(\pm0.0)$ & $\textbf{88.7} ~(\pm0.0)$ & $50.0 ~(\pm0.0)$ & $76.0 ~(\pm0.0)$ & $81.9 ~(\pm0.0)$ & $ 71.3 ~(\pm0.0)$ \\%
        
        FriendBot & FTG & $96.9 ~(\pm 1.1)$ & / & $78.0~(\pm 1.0)$ & / & / & / & / & $75.9 ~(\pm 0.5)$ & - \\%
        
        SATAR & FTG & $93.4~(\pm0.5)$ & / & / & / & / & / & / & $84.0~(\pm0.8)$ & - \\
        
        Botometer & FTG & $57.9$ & $71.6$ & $94.2$ & $89.5$ & $72.6$ & $69.2$ & $50.0$ & $53.1$ & $49.9$ \\
        
        Rodrifuez-Ruiz \textit{et al.} & FTG & $82.4 ~(\pm0.0)$ & /  & $76.4~(\pm0.0)$ & / & / & / & / & $66.0 ~(\pm0.1)$ & $49.4~(\pm0.0)$ \\
        
        GraphHist & FTG & $77.4 ~(\pm 0.2)$ & / & / & / & / & / & / & $51.3 ~(\pm 0.3)$ & - \\
        
        EvolveBot & FTG & $92.2 ~(\pm 1.7)$ & / & / & / & / & / & / & $65.8 ~(\pm 0.6)$ & $71.1 ~(\pm 0.1)$ \\
        
        Dehghan \textit{et al.} & FTG & $62.1 ~(\pm0.0)$ & /  & / & / & / & / & / & $\underline{86.7} ~(\pm0.1)$ & - \\%
        
        GCN & FTG & $96.4~(\pm 0.0 )$ & / & / & / & / & / & / & $77.5~(\pm 0.0)$ & $78.4~(\pm 0.0)$ \\
        
        GAT & FTG & $96.9~(\pm 0.0)$ & / & / & / & / & / & / & $83.3 ~(\pm 0.0)$ & $\underline{79.5}~(\pm 0.0)$ \\
        
        HGT & FTG & $96.0 ~(\pm0.3)$ & / & / & / & / & / & / & $\textbf{86.9} ~(\pm0.2)$ & $74.9 ~(\pm0.1)$ \\
        
        SimpleHGN & FTG & $96.7 ~(\pm0.5)$ & / & / & / & / & / & / & $\underline{86.7} ~(\pm0.2)$ & $76.7 ~(\pm0.3)$ \\
        
        BotRGCN & FTG & $96.5 ~(\pm0.7)$ & /  & / & / & / & / & / & $85.8 ~(\pm0.7)$ & $\textbf{79.7} ~(\pm0.1)$ \\%
        
        RGT & FTG & $97.2 ~(\pm0.3)$ & / & / & / & / & / & / & $86.6 ~(\pm0.4)$ & $76.5 ~(\pm0.4)$ \\%
        
         \bottomrule[1.5pt]
    \end{tabular}
    }
\end{table*}

\subsection{Experiment Results}
We re-implement 35 baseline methods and evaluate them on 9 Twitter bot detection datasets. We run each baseline method for \textbf{five times} and report the average performance and standard deviation. Table \ref{tab:big_acc} presents the benchmarking results. Our main discoveries are summarized as follows:

\begin{itemize}[leftmargin=*]
    \item Graph-based approaches are generally more effective than feature-based or text-based methods.  As a matter of fact, all top 5 models on TwiBot-20 and TwiBot-22 are graph-based. On average, these top-5 graph-based methods outperform the global average of all baselines by 13.8\% and 8.2\% on TwiBot-20 and TwiBot-22. This trend suggests that future research in Twitter bot detection should further examine how users and bots interact on Twitter and the heterogeneous graph structure thus formed.
    \item Most existing datasets do not provide the graph structure of Twitter users to support graph-based approaches, while TwiBot-22 supports all baseline methods and serve as a comprehensive evaluation benchmark. As novel and state-of-the-art models are increasingly graph-based, future Twitter bot detection datasets should provide the graph structure of real-world Twitter.
    \item TwiBot-22 establishes the largest benchmark while exposing the scalability issues of baseline methods. For example, \citet{dehghan2022detectingdehghanetal} achieves near-sota performance on TwiBot-20, while failing to scale to TwiBot-22 as our implementation encounters the out-of-memory problem.
    \item Performance on TwiBot-22 is on average 2.7\% lower than on TwiBot-20 across all baseline methods, which demonstrates that Twitter bot detection is still an open problem that calls for further research. This could be attributed to the fact that Twitter bots are constantly evolving to improve their disguise and evade detection, thus bot detection methods should also adapt and evolve.
\end{itemize}

\subsection{Removing Graphs from Baselines}

 Benchmarking results in Table \ref{tab:big_acc} demonstrate that graph-based approaches generally achieve better performance. To examine the role of graphs in graph-based approaches, we remove the graph component in competitive graph-based methods \citep{Alhosseinietal, moghaddam2022friendshipmoghaddametal, knauth2019languageknauthetal, yang2013empiricalevolvebot, feng2021botrgcn, RGT} and report model performance in Table \ref{tab:remove_graph_all}. It is demonstrated that:
 \begin{itemize}[leftmargin=*]
     \item All baseline methods exhibit performance drops to different extents on two datasets when the graph component is removed. This indicates that the graph-related components in graph-based approaches contribute to bot detection performance.
     \item For graph neural network-based approaches BotRGCN \citep{feng2021botrgcn} and RGT \citep{RGT}, the performance drop is generally more severe. This suggests that graph neural networks play an important role in boosting model performance and advancing bot detection research.
 \end{itemize}
 
 More details about how graphs are removed from baseline methods are presented in Appendix \ref{app:graph_removal}.


\begin{table*}[t]
    \caption{Removing the graph-related model component from graph-based methods (w/o G) while comparing to their original versions (Prev.) on TwiBot-20 and TwiBot-22. \label{tab:remove_graph_all}}
    \vspace{1mm}
    \centering
    \renewcommand{\arraystretch}{1.1}
    \setlength{\tabcolsep}{0.5mm}{
    \resizebox{1\linewidth}{!}{
    \begin{tabular}{c|ccc|ccc|ccc|ccl} 
         \toprule[1.5pt]
         \multirow{3}{*}{\textbf{Method}} & \multicolumn{6}{c|}{\textbf{TwiBot-20}} & \multicolumn{6}{c}{\textbf{TwiBot-22}} \\ \cline{2-13}
         & \multicolumn{3}{c|}{Acc} & \multicolumn{3}{c|}{F1} & \multicolumn{3}{c|}{Acc} & \multicolumn{3}{c}{F1} \\
                          & Prev. & w/o G & Diff. & Prev. & w/o G & Diff. & Prev. & w/o G & Diff. & Prev. & w/o G & Diff. \\  \midrule[1pt]
\citet{Alhosseinietal} & $59.9$ & $61.8$ & $+1.9$ & $72.1$ & $70.7$ & $-1.4$  & $70.7$ & $66.9$ & $-3.8$ & $5.7$  & $3.5$  & $-2.2$  \\
\citet{moghaddam2022friendshipmoghaddametal}  & $74.0$ & $72.2$ & $-1.8$ & $77.9$ & $75.8$ & $-2.1$ & $73.8$ & $73.3$ & $-0.5$ & $32.1$ & $32.0$ & $-0.1$  \\
\citet{knauth2019languageknauthetal}     & $81.9$ & $81.4$ & $-0.5$ & $85.2$ & $84.9$ & $-0.3$ & $71.3$ & $71.5$ & $+0.2$ & $37.1$ & $11.3$ & $-25.8$ \\
EvolveBot \citep{yang2013empiricalevolvebot}                  & $65.8$ & $65.1$ & $-0.7$ & $69.7$ & $69.3$ & $-0.4$ & $71.1$ & $71.0$ & $-0.1$ & $14.1$ & $14.0$ & $-0.1$   \\
BotRGCN \citep{feng2021botrgcn}                    & $85.7$ & $82.6$ & $-3.1$ & $87.3$ & $83.8$ & $-3.5$ & $79.7$ & $75.4$ & $-4.3$ & $57.5$ & $41.2$ & $-16.3$  \\
RGT \citep{RGT}                       & $86.6$ & $82.6$ & $-4.0$ & $88.0$ & $83.8$ & $-4.2$ & $76.5$ & $75.4$ & $-1.1$ & $42.9$ & $41.2$ & $-1.7$ \\ \bottomrule[1.5pt]
    \end{tabular}
    }
    }
\end{table*}

\subsection{Generalization Study}

\begin{figure}[t]
    \centering
    \includegraphics[width=1\textwidth]{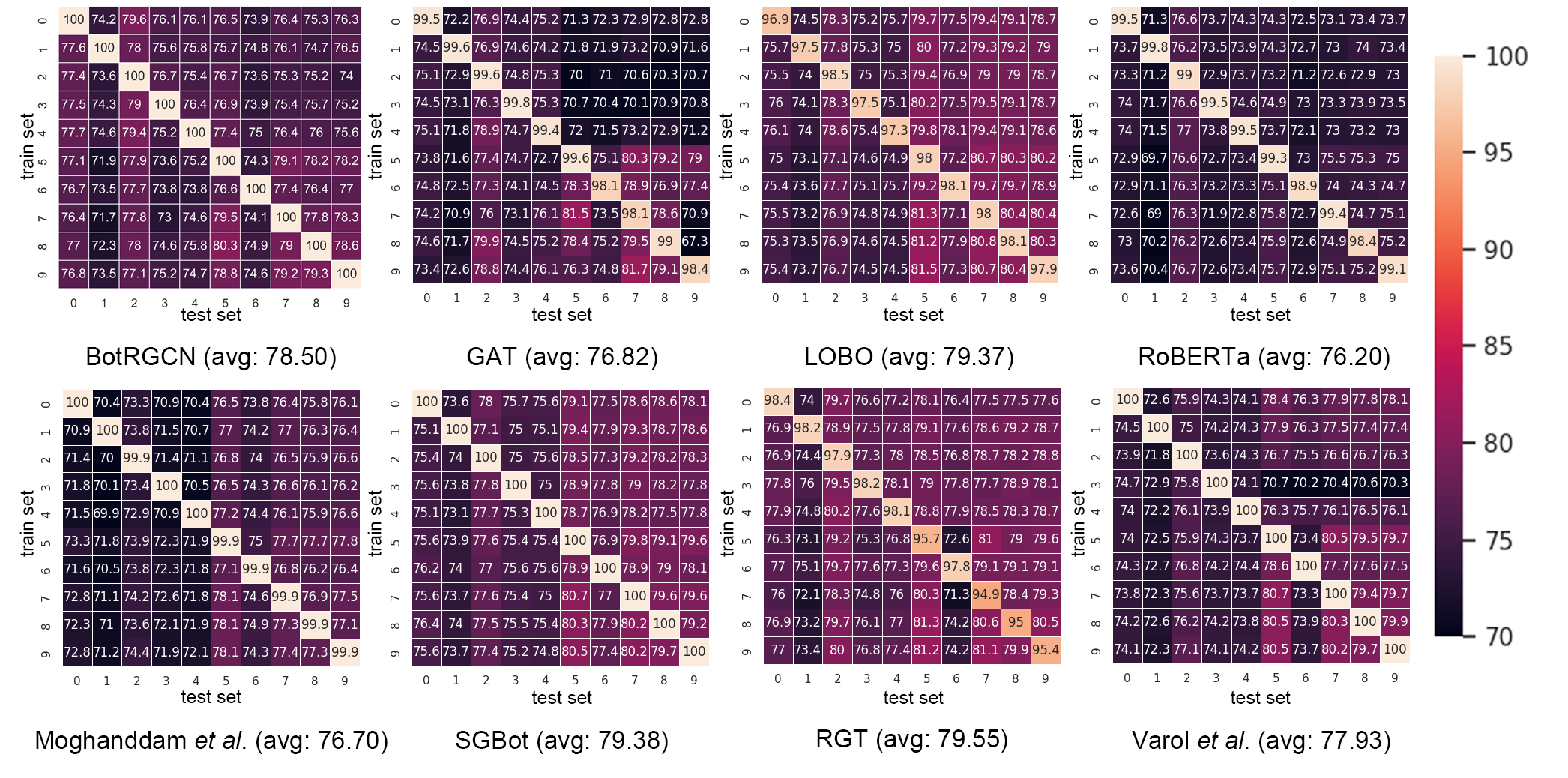}
    \caption{Training models on fold $i$ and testing on fold $j$. We present model accuracy and report the average value of each heatmap (avg), which serves as an overall indicator of generalization ability.}
    \label{fig:heat}
\end{figure}

The challenge of generalization \citep{yang2020scalableyangetal, feng2021satar}, \ie, whether Twitter bot detection models perform well on unseen data, is essential in ensuring that bot detection research translates to effective social media moderation and real-world impact. To evaluate the generalization ability of existing Twitter bot detection approaches, we identify 10 sub-communities in the TwiBot-22 network. We then use these sub-communities as folds and examine the performance of several representative models when trained on fold $i$ and evaluated on fold $j$. Figure \ref{fig:heat} illustrates that:
\begin{itemize}[leftmargin=*]
    \item \textbf{Graph-based methods are better at generalizing to unseen data.} For example, BotRGCN \citep{feng2021botrgcn} achieves the best avg score among all baseline methods, outperforming the second-highest RGT by 3.66. This suggests that leveraging the network structure of Twitter might be a potential solution to the generalization challenge.
    \item \textbf{Good model performance does not necessarily translate to good generalization ability.} For example, GAT outperforms LOBO by 5.9\% and 3.8\% on TwiBot-20 and TwiBot-22 respectively in terms of accuracy. However, GAT has lower avg (-2.55) compared to LOBO. This suggests that future bot detection research should focus on generalization in addition to model performance.
\end{itemize}

More details about the 10 sub-communities are provided in Appendix \ref{app:generalization}.

\section{Evaluation Framework}
We consolidate Twitter bot detection datasets, data prepossessing codes, and all 35 implemented baselines into the TwiBot-22 evaluation framework and make it publicly available. We hope our efforts would facilitate further research in Twitter bot detection through:

\begin{itemize}[leftmargin=*]
    \item establishing a unified interface for different types of Twitter bot detection datasets
    \item providing 35 representative baselines and well-documented implementations
    \item enriching the evaluation framework with new datasets and methods proposed in future research
\end{itemize}

Please refer to \url{https://twibot22.github.io/} for more details.

\section{Conclusion and Future Work}
In this paper, we propose TwiBot-22, a graph-based Twitter bot detection benchmark. TwiBot-22 successfully alleviates the challenges of limited dataset scale, incomplete graph structure, and poor annotation quality in existing datasets. Specifically, we employ a two-stage data collection process and adopt the weak supervision learning strategy for data annotation. We then re-implement 35 representative Twitter bot detection models and evaluate them on 9 datasets, including TwiBot-22, to promote a holistic understanding of research progress. We further examine the role of graphs in graph-based methods and the generalization ability of competitive bot detection baselines. Finally, we consolidate all implemented codes into the TwiBot-22 evaluation framework, where researchers could easily reproduce our experiments and quickly test out new datasets and models.

Armed with the TwiBot-22 benchmark and the TwiBot-22 evaluation framework, we aim to investigate these research questions in the future:
\begin{itemize}[leftmargin=*]
    \item \textbf{How do we identify bot clusters and their coordination campaigns?} While existing works study Twitter bot detection through individual analysis, novel Twitter bots are increasingly observed to act in groups and launch coordinated attacks. We aim to complement the scarce literature by proposing temporal and subgraph-level bot detection approaches to address this issue.
    \item \textbf{How do we incorporate multi-modal user features for bot detection?} In addition to text and graph, Twitter users and bots generate multi-modal user information such as images and videos. Since TwiBot-22 provides user media while none of the 35 baselines leverage these modalities, we aim to further explore Twitter bot detection with the help of images and videos.
    \item \textbf{How do we evaluate the generalization ability of bot detection methods?} Existing works mainly focus on bot detection performance while generalization is essential in ensuring that bot detection research generates real-world impact. We aim to complement the scarce literature by proposing measures to quantitatively evaluate bot detection generalization.
    \item \textbf{How do we improve the scalability of graph-based models?} Existing graph-based bot detection methods demand significantly more computation resources and execution time than feature-based models. Given that the Twitter network is rapidly expanding, we aim to further explore scalable and graph-based bot detection methods.
\end{itemize}

\section*{Acknowledgements}
This work was supported by the National Key Research and Development Program of China (No. 2020AAA0108800), National Nature Science Foundation of China (No. 62192781, No. 62272374, No. 61872287, No. 62250009, No. 62137002), Innovative Research Group of the National Natural Science Foundation of China (61721002), Innovation Research Team of Ministry of Education (IRT\_17R86), Project of China Knowledge Center for Engineering Science and Technology and Project of Chinese academy of engineering ``The Online and Offline Mixed Educational Service System for ‘The Belt and Road’ Training in MOOC China''.

We would like to thank the reviewers and area chair for the constructive feedback. We would also like to thank all LUD lab members for our collaborative research environment and for making our bot detection research series possible. As Shangbin concludes his term as the director and Zhaoxuan begins his term, we hope the LUD lab will continue to thrive in the years to come.

\bibliography{neurips_data_2021}
\bibliographystyle{unsrtnat}

\clearpage

\section*{Checklist}

\begin{enumerate}
    \item For all authors...
    \begin{enumerate}[label=(\alph*)]
        \item Do the main claims made in the abstract and introduction accurately reflect the paper's contributions and scope? {\color{blue} [Yes]}
        \item Have you read the ethics review guidelines and ensured that your paper conforms to them? {\color{blue} [Yes]}
        \item Did you discuss any potential negative societal impacts of your work? {\color{blue} [Yes]} Section A.6
        \item Did you describe the limitations of your work? {\color{blue} [Yes]} Section A.6
    \end{enumerate}
    \item If you are including theoretical results...
    \begin{enumerate}[label=(\alph*)]
        \item Did you state the full set of assumptions of all theoretical results? {\color{gray} [N/A]}
        \item Did you include complete proofs of all theoretical results? {\color{gray} [N/A]}
    \end{enumerate}
    \item If you ran experiments...
    \begin{enumerate}[label=(\alph*)]
        \item Did you include the code, data, and instructions needed to reproduce the main experimental results (either in the supplemental material or as a URL)? {\color{blue} [Yes]} Please refer to the TwiBot-22 GitHub repository listed in Section A.6.
        \item Did you specify all the training details (e.g., data splits, hyperparameters, how they were chosen)? {\color{blue} [Yes]} Please refer to the TwiBot-22 GitHub repository listed in Section A.6.
        \item Did you report error bars (e.g., with respect to the random seed after running experiments multiple times)? {\color{blue} [Yes]}
        \item Did you include the amount of compute and the type of resources used (e.g., type of GPUs, internal cluster, or cloud provider)? {\color{blue} [Yes]} Section B.6
    \end{enumerate}
    \item If you are using existing assets (e.g., code, data, models) or curating/releasing new assets...
    \begin{enumerate}[label=(\alph*)]
        \item If your work uses existing assets, did you cite the creators? {\color{blue} [Yes]}
        \item Did you mention the license of the assets? {\color{blue} [Yes]} Section A.6
        \item Did you include any new assets either in the supplemental material or as a URL? {\color{blue} [Yes]} We provide the TwiBot-22 dataset as a new asset and provide URLs in Section A.6.
        \item Did you discuss whether and how consent was obtained from people whose data you're using/curating? {\color{blue} [Yes]} We strictly follow the original license of existing datasets and rules in the Twitter Developer Agreement and Policy.
        \item Did you discuss whether the data you are using/curating contains personally identifiable information or offensive content? {\color{blue} [Yes]} Our dataset does not contain the information of private and protected users. Since TwiBot-22 aims to facilitate bot detection research and certain bots are designed to be offensive, there might be offensive content.
    \end{enumerate}
    \item If you used crowdsourcing or conducted research with human subjects...
    \begin{enumerate}[label=(\alph*)]
        \item Did you include the full text of instructions given to participants and screenshots, if applicable? {\color{gray} [N/A]}
        \item Did you describe any potential participant risks, with links to Institutional Review Board (IRB) approvals, if applicable? {\color{gray} [N/A]}
        \item Did you include the estimated hourly wage paid to participants and the total amount spent on participant compensation? {\color{gray} [N/A]}
    \end{enumerate}
\end{enumerate}

\clearpage

\appendix

\section{TwiBot-22 Details}

\subsection{Entities and Relations}
TwiBot-22 collects four types of entities on the Twitter social network: user, tweet, list, and hashtag. The detailed information of these entities is shown in Table~\ref{tab:entities} while a complete list of all relation types in TwiBot-22 is presented in Table \ref{tab:relation_types}.

\begin{table*}[]
    \centering
    \caption{Entities in the TwiBot-22 heterogeneous graph.}
    \begin{tabular}{c|c|c}
        \toprule[1.5pt]
        \textbf{Entity Name}&\textbf{Description}&\textbf{Main Metadata}\\
        \midrule[1pt]
        \multirow{5}{*}{User}&\multirowcell{5}[0pt][l]{Users are the most important \\entity on Twittersphere.}&\multirowcell{5}{created at, description, entities\\location, name, profile image url\\protected, url, username\\verified, withheld, followers count\\following count, tweet count, listed count}\\
        &&\\
        &&\\
        &&\\
        &&\\
        \hline
        \multirow{5}{*}{Tweet}&\multirowcell{5}[0pt][l]{Users post tweets to share their\\ thoughts and interact with other\\ users.}&\multirowcell{5}{attachments, context annotations, entities\\created at, geo, lang, possibly sensitive\\referenced tweets, reply settings\\source, text, withheld, retweet count\\reply count, like count, quote count}\\
        &&\\
        &&\\
        &&\\
        &&\\
        \hline
        \multirow{4}{*}{List}&\multirowcell{4}[0pt][l]{A list is curated feeds from\\ selected users that allow you \\to listen to relevant discussions\\ or influencers.}&\multirowcell{4}{private, created at, description\\name, follower count, member count}\\
        &&\\
        &&\\
        &&\\
        \hline
        \multirow{4}{*}{Hashtag}&\multirowcell{4}[0pt][l]{A hashtag is a metadata tag that \\is prefaced by "\#".\\ It is used to link tweets with the\\ same theme together.}&\multirow{4}{*}{hashtag name}\\
        &&\\
        &&\\
        &&\\
        \bottomrule[1.5pt]
    \end{tabular}
    
    \label{tab:entities}
\end{table*}

\begin{table}[t]
    \centering
    \caption{Relations in the TwiBot-22 heterogeneous graph.}
    \begin{tabular}{c|c|c|c}
        \toprule[1.5pt]
         \textbf{Relation}&\textbf{Source Entity}&\textbf{Target Entity}&\textbf{Description}\\ \midrule[1pt]
         following&user&user& user A follows user B\\
         followers&user&user& user A is followed by user B\\
         post&user&tweet& user A posts tweet B\\
         pinned&user&tweet& user A pins tweet B\\
         like&user&tweet& user A likes tweet B\\
         mentioned&tweet&user& tweet A mentions user B\\
         retweeted&tweet&tweet& tweet A retweets tweet B\\
         quoted&tweet&tweet& tweet A quotes tweet B with comments\\
         reply\_to&tweet&tweet& tweet A replies to tweet B\\
         own&user&list& user A is the creator of list B\\
         membership&list&user& user A is a member of list B\\
         followed&list&user& user A follows list B\\
         contain&list&tweet& list A contains tweet B\\
         discuss&tweet&hashtag &tweet A discussed hashtag B\\ \bottomrule[1.5pt]
    \end{tabular}
    
    \label{tab:relation_types}
\end{table}

\subsection{Data Collection Details}
\label{app:collection_detail}
\begin{table}[t]
    \centering
    \caption{User metadata adopted in diversity-aware sampling.}
    \begin{tabular}{c|c|c}
        \toprule[1.5pt]
        \textbf{Metadata Name}& \textbf{Description}&\textbf{Type}\\
        \midrule[1pt]
        active days&days between user creation time and collected time&numerical\\
        following count&number of user followings&numerical\\
        followers count&number of user followers&numerical\\
        tweet count&number of user tweets&numerical\\
        listed count&number of user lists&numerical\\
        verified&whether the user is verified or not &true-or-false\\
        homepage url&whether user has urls in homepage or not&true-or-false\\
        \bottomrule[1.5pt]
    \end{tabular}
    \label{tab:metadata}
\end{table}

\paragraph{The complete process of data collection.} For the first stage of user network collection, we adopt \textit{@NeurIPSConf} as the starting user. We use the Twitter API to retrieve 1,000 followers and 1,000 followees as the user's neighborhood for BFS expansion. We randomly adopt one of the two sampling strategies (distribution diversity or value diversity) and randomly select one metadata from Table~\ref{tab:metadata} to include 6 users from its neighborhood into the TwiBot-22 dataset. We then randomly select one unexpanded user in TwiBot-22 for a new round of neighborhood expansion. For the second stage of heterogeneous graph building, we first collect 1,000 tweets for the user in the user network, and 200 tweets for the user expanded from the user network. We collect the pinned tweet and the recent 100 liked tweet of each user in TwiBot-22. For each tweet we collect now, we collect the tweets it retweets, quotes, or replies and the users it mentions. We collect a user's recent 100 lists with the newest 100 members, followers, and tweets. We collect all hashtags in the tweets and search for more tweets related to a hashtag using Twitter API. Finally, we make sure that the creator of each tweet is collected and collect 40 tweets of these users according to \citet{ng2022stabilizing} for stable benchmarking

\paragraph{Data collection time.} The first stage of user network collection is conducted from January 20th, 2022 to February 1st, 2022. The second stage of heterogeneous graph building is conducted from February 1st, 2022 to March 15th, 2022.

\subsection{Expert Annotation Details}
\label{app:expert_annotation}
We invite 17 researchers in our group who are active Twitter users, are familiar with bot detection literature, and have conducted experiments with the TwiBot-20 datasets. We then assign each Twitter user in TwiBot-22 to 5 different experts and ask them to evaluate whether the user is a human, a bot, or not sure. We use majority voting to obtain the expert annotations of these 1,000 users, which are then leveraged to guide the weak supervision leaning process.

\subsection{Dataset Statistics} 
\label{app:dataset_stat}
Table~\ref{tab:statistic} presents important statistics about the TwiBot-22 benchmark.
\begin{table}[t]
    \centering
    \caption{Statistics of TwiBot-22.}
    \begin{tabular}{c|c|c|c|c|c}
    \toprule[1.5pt]
    \textbf{Item}&\textbf{Value}&\textbf{Item}&\textbf{Value}&\textbf{Item}&\textbf{Value}\\
    \midrule[1pt]
    entity type&4&post&88,217,457&following&2,626,979\\
    relation type&14&pin&347,131&follower&1,116,655\\
    user&1,000,000&like&595,794&contain&1,998,788\\
    hashtag&5,146,289&mention&4,759,388&discuss&66,000,633\\
    list&21,870&retweet&1,580,643&bot&139,943\\
    tweet&88,217,457&quote&289,476&human&860,057\\
    user metadata&17&reply&1,114,980&entity&92,932,326\\
    hashtag metadata&2&own&21,870&relation&170,185,937\\
    list metadata&8&member&1,022,587&max degree&270,344\\
    tweet metadata&20&follow&493,556&verified user&95,398\\
    \bottomrule[1.5pt]
    \end{tabular}
    
    \label{tab:statistic}
\end{table}

\subsection{Annotation Quality Study Details}
\label{app:annotation_quality}
To compare the annotation quality of TwiBot-22 and TwiBot-20, we ask 6 researchers to participate in an expert study. They are familiar with Twitter bot detection research and most of them have previously published on this topic. Specifically, we randomly select 500 users from TwiBot-20 and TwiBot-22 respectively and assign each user to 3 experts. We then ask them to evaluate each user as "definitely bot", "likely bot", "not sure", "likely human", and "definitely human". Based on their evaluations, we calculate the accuracy and F1-score between expert opinions and dataset labels. We also report the Randolph's Kappa Coefficient \citep{randolph2005free}, which models agreement between experts in Figure \ref{fig:data_analysis}(c).

\subsection{Other TwiBot-22 Details}

\paragraph{Dataset documentation.} We encourage the readers to refer to the TwiBot-22 evaluation framework (\url{https://github.com/LuoUndergradXJTU/TwiBot-22}) for the documentation of TwiBot-22 and the TwiBot-22 evaluation framework.

\paragraph{Intended use.} TwiBot-22 should be used for research in Twitter bot detection and social network analysis.

\paragraph{Relevent URLs.} We list all TwiBot-22 URLs in the following.

\begin{itemize}[leftmargin=*]
    \item \textbf{Official TwiBot-22 website} (\url{https://twibot22.github.io/}) is the main reference of TwiBot-22, presenting our dataset, paper, evaluation framework, and contact information.
    \item \textbf{TwiBot-22 repository} (\url{https://github.com/LuoUndergradXJTU/TwiBot-22}) hosts implemented codes for dataset preprocessing and 35 Twitter bot detection methods. The repository is well documented to facilitate reproducible research.
    \item \textbf{TwiBot-22 dataset} will be permanently hosted on our group Google Drive account (\url{https://drive.google.com/drive/folders/1YwiOUwtl8pCd2GD97Q_WEzwEUtSPoxFs?usp=sharing}).
\end{itemize}

\paragraph{Hosting and maintenance.} The TwiBot-22 dataset will be hosted via Google Drive and regularly maintained by our research group. The TwiBot-22 evaluation framework will be hosted on GitHub. Our team will continue to add new datasets and baselines with the help of the research community.

\paragraph{Licensing.} The TwiBot-22 dataset uses the CC BY-NC-ND 4.0 license. Implemented code in the TwiBot-22 evaluation framework uses the MIT license.

\paragraph{Author statement.} We bear all responsibility in case of violation of rights, etc., and confirmation of the data license.

\paragraph{Limitations.} One minor limitation of TwiBot-22 is that we do not download and store user media (images and videos) in TwiBot-22, while these multimedia content might be useful for bot detection. However, if researchers do deem multimedia content as necessary for bot detection, they can download with media links in TwiBot-22 by themselves.

\paragraph{Potential negative societal impact.} Although TwiBot-22 and the TwiBot-22 evaluation framework are designed to facilitate bot detection research and improve bot detection models, it might be abused by bot operators to examine the characteristics of bots that evade detection, and thus designing bot algorithms that are more evasive. We need to make sure that the TwiBot-22 dataset and evaluation framework should not be abused to design advanced Twitter bots.

\section{Experiment Details}

\subsection{Baseline Details}
\label{app:baseline_detail}
We briefly describe each of the 35 Twitter bot detection baseline methods:
\begin{itemize}[leftmargin=*]
    \item \textbf{SGBot} \citep{yang2020scalableyangetal}. SGBot is proposed to address the scalability and generalization problem in Twitter bot detection. SGBot leverages 8 types of user metadata such as status count and 12 derived features such as tweet frequency and adopts random forest to identify bot.
    \item \textbf{Kudugunta \textit{et al.}} \citep{kudugunta2018deepkuduguntaetal}. The baseline addresses two challenges, account-level classification and tweet-level classification. In the task of account-level classification, the baseline introduces a technique that combines synthetic minority oversampling (SMOTE) with undersampling techniques. In the task of tweet-level classification, the baseline introduces an architecture called contextual LSTM.
    \item \textbf{Hayawi \textit{et al.}}  \citep{hayawi2022deeprobothayawietal}. DeeProBot, which is short for Deep Profile-based Bot detection Framework, utilizes different types of features including user's numerical or binary metadata and user's description, making the model more comprehensive. DeeProBot uses GLoVe word embeddings to get the embedding of the textual information. LSTM and dense layers are then used to learn user representations for bot detection.
    \item \textbf{BotHunter} \citep{beskow2018botbothunter}. In this work, the features are composed of user attributes, network attributes, contents, and timing information. After extracting the above features, random forest is exploited as the classifier.
    \item \textbf{NameBot} \citep{beskow2019itsbeskowetalname}. NameBot utilizes twitter username as the only classification basis and extracts numerical features such as TF-IDF. The extracted numeric features are then exploited as input of a linear regression classifier. While achieving great good accuracy on training set, this basline shows poor transferbility on new datasets.
    \item \textbf{Abreu \textit{et al.}} \citep{abreu2020twitterabreuetal}. This model chooses five essential Twitter user features to conduct relative experiments. They calculated accuracy, AUC, recall and F1-score on several existing datasets with four machine learning algorithms.
    \item \textbf{Cresci \textit{et al.}} \citep{cresci2016dna1}. This method encodes different action types with different characters, thus representing usernames by strings. Users who share the longest common substring are considers as bots have similar behaviors.
    \item \textbf{Wei \textit{et al.}} \citep{wei2019twitterweietal}. This paper use pre-trained GloVe word vectors on Twitter as word embedding. Multiple layers of bidirectional LSTMs are used for bot detection.
    \item \textbf{BGSRD} \citep{guo2021socialbgsrd}. This model utilizes BERT and GCN to achieve social bots detection. In specific, this paper use word and user description as graph nodes, there are no connection within users and words. In training steps, BGSRD first uses BERT (Roberta) to process users' description, the output of BERT layer is the initial feature of graph nodes, and the initial feature of words node is zeros. The final prediction is the combination of BERT and GAT output.
    \item \textbf{RoBERTa} \citep{liu2019roberta}. This baseline leverages pre-trained language model RoBERTa to encode user tweets and descriptions, then feed them to an MLP to distinguish bots from human.
    \item \textbf{T5} \citep{raffel2019exploringt5}. In this baseline, we use pretained language model T5-small to encode tweets and descriptions, then feed them into an MLP classifier.
    \item \textbf{Efthimion \textit{et al.}} \citep{efthimion2018supervisedefthimionetal}. This paper leverages a wide range of users' features including length of user names, reposting rate, temporal patterns, sentiment expression, followers-to-friends ratio, and message variability for bot detection. Logistic regression and support vector machine are applied successively for profile and account activity analysis. Levenshtein distance is applied for text mining.
    \item \textbf{Kantepe \textit{et al.}} \citep{kantepe2017preprocessingkantepeetal}. This paper explores the importance of features by comparing sixty-two different features related to Twitter account properties and tweet contents, and selects the most important eleven features through experimental comparison. The classification task was then performed using 62 or 11 features as input with classifier like GBDT or SVM.
    \item \textbf{Miller \textit{et al.}} \citep{miller2014twittermilleretal}. Miller et al. proposed a clustering based method to detect spam accounts in twitter. Specifically, they exploit classic clustering algorithms such as DBSCAN and K-MEANS on human accounts to obtain human clusters. In test phase, users whose clusters can not be filed under any of the existing human clusters are consider as bots.
    \item \textbf{Varol \textit{et al.}} \citep{varol2017onlinevaroletal}. User metadata, tweet content, friends, sentiment and network statistics are adapted as user's features, then the extracted features are utilized as inputs of a random forest model.
    \item \textbf{Kouvela \textit{et al.}} \citep{kouvela2020botbotdetective}. This baseline leverages user features and content features from each user and classifies users with random forest. Specifically, it uses 36 features from each account and the content features are extracted from the latest 20 tweets.
    \item \textbf{Santos \textit{et al.}} \citep{ferreira2019uncoveringsantosetal}. This baseline extracts 16 features from users' tweets and descriptions and feed the features into a decision tree for classification. 
    \item \textbf{Lee \textit{et al.}} \citep{lee2011sevenleeetal}. This method introduces the social honeypots to attract bot users by manipulating the honeypot users' tweets frequency and social network structure. After analyzing the data collected by social honeypots, 18 features are selected and fed into the random forest classifier. 
    \item \textbf{LOBO} \citep{echeverri2018lobo}. This baseline extracts 19 features (26 on Twibot-22) from each user and adopts random forest for classification.
    \item \textbf{Moghaddam \textit{et al.}} \citep{moghaddam2022friendshipmoghaddametal}. This model combines profile-based features and friendship preference features, which compares the distribution of followers' features and sub-population of accounts. 
    \item \textbf{Alhosseini \textit{et al.}} \citep{Alhosseinietal}. This model uses age, statuses\_count, account length name, followers\_count, friends\_count and favourites\_count as user features and feed them into a GCN layer to identify bot users.

    \item \textbf{Knauth \textit{et al.}} \citep{knauth2019languageknauthetal}. This paper extracts features from user's meta data, tweets, user behavior, and feeds these features into Adaboost classifier.
    \item \textbf{FriendBot} \citep{beskow2020youbeskowetalfriend}. This paper introduces network metrics into twitter bot detection tasks. Specifically, they construct a 2-hop ego network for each twitter account based on four types of relations: following, retweet, mention, and reply. They collect account features based on metrics of these ego networks. Finally, they exploit random forest algorithm for classification.
    \item \textbf{SATAR} \citep{feng2021satar}. SATAR is a self-supervised representation learning framework of Twitter users. SATAR jointly uses semantic, property, and neighborhood information and adopts a co-influence module to aggregate these information.  SATAR considers the follower count as self-supervised label to pre-train parameters and fine-tune parameters in bot detection task.
    \item \textbf{Botometer} \citep{yang2022botometer}. Botometer (formerly BotOrNot) is a public website to check the activity of a Twitter account and give it a score, where higher scores mean more bot-like activity. Botometer's classification system leverages more than 1,000 features using available meta-data and information extracted from interaction patterns and content.
    \item \textbf{Rodriguez-Ruiz \textit{et al.}} \citep{rodriguez2020onerodriguezetal}. This paper designs a one-class classification model, which uses the social network and tweet information of Twitter users to extract 13 features for feature engineering, and the model has also achieved good classification results.
    \item \textbf{GraphHist} \citep{GraphHist}. The authors design a new graph classifier based on histogram and customized backward operator. By exploiting the ego-graph of twitter users, bot detection can be solved by utilizing the proposed graph-level classifier.
    \item \textbf{EvolveBot} \citep{yang2013empiricalevolvebot}. This method designs 11 robust features, together with 7 efficient features to combat evasion tatics of spammers.
    \item \textbf{Dehghan \textit{et al.}} \citep{dehghan2022detectingdehghanetal}. This baseline combines the profile features, text features, and graph features for bot detection. After obtaining account representations through Deepwalk and struc2vec, XGBclassifier is applied to identify bot users.
    \item \textbf{GCN} \citep{kipf2016semigcn}. GCN aggregates features from neighbors equally and learns representation for each user. These representations are passed to an MLP for classification. The initial user features are identical with BotRGCN.
    \item \textbf{GAT} \citep{velivckovic2017graphgat}. Graph Attention Network (GAT) introduces attention mechanism to GNN models, making it capable of distinguishing the importance of neighboring users in aggregation. Same as GCN, it can learn user representations and feed them into an MLP for classification. The initial user features are identical with BotRGCN.
    \item \textbf{HGT} \citep{hu2020heterogeneoushgt}. Heterogeneous Graph Transformers (HGT) is a dedicated heterogeneous GNN that mainly consists of two modules, heterogeneous mutual attention and heterogeneous message passing. Heterogeneous mutual attention considers the edge type and source and target node type when calculating attention scores. Heterogeneous message passing module incorporates the source node type and the edge dependency in passed messages. The initial user features are identical with BotRGCN.
    \item \textbf{simpleHGN} \citep{lv2021wesimplehgn}. SimpleHGN is a simple yet effective GNN for heterogeneous graph inspired by the GAT. SimpleHGN adopts three strategies to enhance GAT, learnable embedding for each edge-type, node-level and edge-level residual connections as well as the L2 regularization on output representations. The initial user features are identical with BotRGCN.
    \item \textbf{BotRGCN} \citep{feng2021botrgcn}. BotRGCN utilizes the text information from user descriptions and tweets, as well as numerical and categorical user property information. Then BotRGCN constructs a heterogeneous graph from the Twitter network based on user relationships and relational graph convolutional networks (R-GCN) is applied to learn user representations for bot detection tasks.
    \item \textbf{RGT} \citep{RGT}. Relational Graph Transformers is a GNN framework that uses graph transformers and semantic attention network to model the intrinsic influence heterogeneity and relation heterogeneity in Twittersphere.  Specifically, RGT first learns users' representation under each relation with graph transformers, then it aggregate representations from all relations using the semantic attention network. 
\end{itemize}

\begin{table*}[]
\renewcommand\arraystretch{1.3}
    \caption{Average model performance (F1-score) and standard deviation of 35 baseline methods on 9 datasets. \textbf{Bold} and \underline{underline} indicate the highest and second highest performance. The F, T, and G in the "Type" column stands for feature, text, and graph. Cresci et al. and Botometer are deterministic methods without standard deviation. / indicates that the dataset could not support the baseline. - indicates that the baseline could not scale to the largest TwiBot-22 dataset.}
    \centering
    \resizebox{1\linewidth}{!}{
    \begin{tabular}{ccccccccccc}
        \toprule[1.5pt]
        \textbf{Method} & \textbf{Type} & \textbf{C-15} & \textbf{G-17} & \textbf{C-17} & \textbf{M-18} & \textbf{C-S-18} & \textbf{C-R-19} & \textbf{B-F-19} & \textbf{TwiBot-20} & \textbf{TwiBot-22} \\ 
        \midrule[1pt]
        
        SGBot & F & $77. 9~(\pm0.1)$ & $\underline{72.1} ~(\pm1.2)$ & $94.6 ~(\pm0.2)$ & $\underline{99.5} ~(\pm0.0)$ & $82.3 ~(\pm0.1)$ & $82.7 ~(\pm1.7)$ & $49.6 ~(\pm3.4)$ & $84.9 ~(\pm0.4)$ & $36.6 ~(\pm0.2)$ \\%
        
        Kudugunta \textit{et al.} & F &$75.3 ~(\pm 0.2)$ & $49.8 ~(\pm 2.1)$ & $91.7 ~(\pm 0.2)$ & $94.5 ~(\pm 0.3)$ & $50.9 ~(\pm 0.4)$ & $49.2 ~(\pm 1.3)$ & $49.6 ~(\pm 8.2)$ & $47.3 ~(\pm 1.4)$ & $51.7 ~(\pm 0.0)$ \\%
        
        Hayawi \textit{et al.} & F & $85.6 ~(\pm0.0)$ & $34.7 ~(\pm0.1)$  & $93.8 ~(\pm0.0)$ & $91.5 ~(\pm0.0)$ & $60.8 ~(\pm0.1)$ & $60.9 ~(\pm0.0)$ & $20.5 ~(\pm0.1)$ & $77.1 ~(\pm0.0)$ & $24.7 ~(\pm0.1)$ \\%
        
        BotHunter & F & $97.2 ~(\pm1.0)$ & $69.2 ~(\pm1.0)$ & $91.6 ~(\pm0.1)$ & $\textbf{99.6} ~(\pm0.0)$ & $82.2 ~(\pm0.2)$ & $\underline{82.9} ~(\pm1.9)$ & $49.6 ~(\pm3.1)$ & $79.1~(\pm0.4)$ & $23.5 ~(\pm0.1)$ \\
        
        NameBot & F & $83.4 ~(\pm0.0)$ & $44.8 ~(\pm0.0)$  & $85.7 ~(\pm0.0)$ & $91.6 ~(\pm0.0)$ & $61.1 ~(\pm0.0)$ & $67.5 ~(\pm0.0)$ & $38.5 ~(\pm0.0)$ & $65.1 ~(\pm0.1)$ & $0.5~(\pm0.0)$\\
        
        Abreu \textit{et al.} & F & $76.4 ~(\pm0.1)$ & $66.7 ~(\pm0.1)$ & $95.0~(\pm0.1)$ & $97.9 ~(\pm0.1)$ & $76.9 ~(\pm0.1)$ & $\textbf{83.5} ~(\pm0.1)$ & $\textbf{53.8} ~(\pm0.1)$ & $77.1~(\pm0.1)$ & $53.4 ~(\pm0.1)$ \\
        \midrule[1pt]
        
        Cresci \textit{et al.} & T & $1.17$ & / & $22.8$ & / & / & / & / & $13.7$ & - \\
        
        Wei \textit{et al.} & T & $82.7 ~(\pm2.2)$& /& $78.4 ~(\pm1.7)$ & /& /& / &/ & $57.3 ~(\pm3.1)$ & $53.6 ~(\pm1.3)$ \\%
        
        BGSRD & T & $90.8 ~(\pm0.6)$ & $35.7 ~(\pm32.6)$ & $86.3 ~(\pm0.0)$ & $90.5 ~(\pm1.0)$ & $58.2 ~(\pm12.0)$ & $41.1 ~(\pm13.0)$ & $13.0 ~(\pm13.0)$ & $70.0 ~(\pm2.6)$ & $21.1 ~(\pm 29.0)$\\
        
        RoBERTa & T & $95.8 ~(\pm 0.1) $ & / & $94.3 ~(\pm 0.1)$ & / & / & / & / & $73.1 ~(\pm 0.5)$ & $20.5 ~(\pm 1.7)$ \\
        
        
        T5 & T & $89.3 ~(\pm 0.2)$ & / & $92.3 ~(\pm 0.1)$ & / & / & / & / & $70.5 ~(\pm 0.3)$ & $20.2 ~(\pm 2.0)$\\%
        \midrule[1pt]

        Efthimion \textit{et al.} & FT & $94.1 ~(\pm 0.0)$ & $5.2 ~(\pm 0.0)$ & $91.8 ~(\pm 0.0)$ & $95.9 ~(\pm 0.0)$ & $68.2 ~(\pm 0.0)$ & $71.7~(\pm 0.0)$ & $0.0~(\pm 0.0)$ & $67.2 ~(\pm 0.0)$ & $27.5~(\pm 0.0)$ \\%
        
        Kantepe \textit{et al.} & FT & $78.2 ~(\pm1.4)$ & / & $79.4 ~(\pm1.3)$ & / & / & /&/ & $62.2 ~(\pm2.1)$ & $\textbf{58.7} ~(\pm1.6)$ \\%
        
        Miller \textit{et al.} & FT & $83.8 ~(\pm0.0)$ & $59.9 ~(\pm0.0)$ & $86.8 ~(\pm0.1)$ & $91.1 ~(\pm0.0)$ & $56.8 ~(\pm0.0)$ & $43.6 ~(\pm0.0)$ & $0.0 ~(\pm0.0)$ & $74.8 ~(\pm0.3)$ & $45.3 ~(\pm0.0)$ \\

        Varol \textit{et al.} & FT & $94.7 ~(\pm0.4)$ & / & / & / & / & / & / & $81.1 ~(\pm0.5)$ & $27.5 ~(\pm0.3)$ \\
        
        Kouvela \textit{et al.} & FT & $98.2 ~(\pm0.4)$ & $66.6 ~(\pm1.7)$ & $\underline{99.1} ~(\pm0.1)$ & $98.2 ~(\pm0.1)$ & $80.4 ~(\pm0.2)$ & $81.1 ~(\pm1.0)$ & $28.1 ~(\pm5.3)$ & $86.5 ~(\pm0.3)$ & $30.0 ~(\pm0.0)$ \\%
        
        Santos \textit{et al.} & FT & $78.8 ~(\pm 0.0)$ & $14.5 ~(\pm 0.0)$ & $83.0 ~(\pm 0.0)$ & $92.4 ~(\pm 0.0)$ & $65.2 ~(\pm 0.0)$ & $75.7 ~(\pm 0.0)$ & $21.0 ~(\pm 0.0)$ & $60.3 ~(\pm 0.0)$ & - \\
        
        Lee \textit{et al.} & FT & $\underline{98.6} ~(\pm0.1)$ & $67.8 ~(\pm1.8)$ & $\textbf{99.3} ~(\pm0.0)$ & $97.9 ~(\pm0.1)$ & $\underline{82.5} ~(\pm0.4)$ & $82.7 ~(\pm1.8)$ & $\underline{50.3} ~(\pm3.2)$ & $80.0 ~(\pm0.5)$ & $30.4 ~(\pm0.2)$ \\%
        
        LOBO & FT & $\textbf{98.8} ~(\pm0.3)$ & / & $97.7 ~(\pm0.2)$ & / & / & / & / & $80.8 ~(\pm0.2)$ & $ 38.6 ~(\pm0.2)$ \\
        \midrule[1pt]

        Moghaddam \textit{et al.} & FG & $73.9 ~(\pm0.2)$ & / & / & / & / & / & / & $79.9 ~(\pm0.7)$ & $32.1 ~(\pm0.0)$ \\
        
        Alhosseini \textit{et al.} & FG & $92.2 ~(\pm 0.4)$ & / & / & / & / & / & / & $72.0 ~(\pm 0.5)$ & $38.1 ~(\pm 5.9)$ \\
        \midrule[1pt]
        
        Knauth \textit{et al.} & FTG & $91.2 ~(\pm0.0)$ & $39.1 ~(\pm0.0)$ & $93.4 ~(\pm0.0)$ & $91.3 ~(\pm0.0)$ & $\textbf{94.0} ~(\pm0.0)$ & $54.2 ~(\pm0.0)$ & $41.3 ~(\pm0.0)$ & $85.2 ~(\pm0.0)$ & $ 37.1 ~(\pm0.0)$ \\%
        
        FriendBot & FTG & $97.6 ~(\pm 0.8)$ & / & $87.4~(\pm 0.5)$ & / & / & / & / & $80.0 ~(\pm 0.3)$ & - \\%
        
        SATAR & FTG & $95.0~(\pm0.3)$ & / & / & / & / & / & / & $86.1~(\pm0.7)$ & - \\
        
        Botometer & FTG & $66.9$ & $\textbf{77.4}$ & $96.1$ & $46.0$ & $79.6$ & $79.0$ & $30.8$ & $53.1$ & $42.8$ \\
        
        Rodrifuez-Ruiz \textit{et al.} & FTG & $87.7~(\pm0.0)$ & /  & $85.7~(\pm0.0)$ & / & / & / & / & $63.1 ~(\pm0.1)$ & $56.6~(\pm0.0)$ \\
        
        GraphHist & FTG & $84.5 ~(\pm 8.2)$ & / & / & / & / & / & / & $67.6 ~(\pm 0.3)$ & - \\
        
        EvolveBot & FTG & $90.1 ~(\pm 2.0)$ & / & / & / & / & / & / & $69.7 ~(\pm 0.5)$ & $14.1 ~(\pm 0.1)$ \\
        
        Dehghan \textit{et al.} & FTG & $88.3 ~(\pm0.0)$ & /  & / & / & / & / & / & $76.2 ~(\pm0.0)$ & - \\%
        
        GCN & FTG & $97.2~(\pm 0.0)$ & / & / & / & / & / & / & $80.8~(\pm 0.0)$ & $54.9~(\pm 0.0)$ \\
        
        GAT & FTG & $97.6~(\pm 0.0)$ & / & / & / & / & / & / & $85.2 ~(\pm 0.0)$ & $55.8~(\pm 0.0)$ \\
        
        HGT & FTG & $96.9 ~(\pm0.2)$ & / & / & / & / & / & / & $\textbf{88.2} ~(\pm0.2)$ & $39.6 ~(\pm2.1)$ \\
        
        SimpleHGN & FTG & $97.5 ~(\pm0.4)$ & / & / & / & / & / & / & $\textbf{88.2} ~(\pm0.2)$ & $45.4 ~(\pm0.4)$ \\
        
        BotRGCN & FTG & $97.3 ~(\pm0.5)$ & /  & / & / & / & / & / & $87.3 ~(\pm0.7)$ & $\underline{57.5} ~(\pm1.4)$ \\%
        
        RGT & FTG & $97.8 ~(\pm0.2)$ & / & / & / & / & / & / & $\underline{88.0} ~(\pm0.4)$ & $42.9 ~(\pm0.5)$ \\%
        
         \bottomrule[1.5pt]
    \end{tabular}
    }

    \label{tab:big_f1}
\end{table*}

\subsection{F, T, or G?}
We categorize baseline methods into F, T, or G with the following rules:
\begin{itemize}[leftmargin=*]
    \item If the baseline leverages user metadata and conduct feature engineering, the baseline is F.
    \item If the baseline leverages the content of tweets and user description texts, the baseline is T.
    \item If the baseline leverages the network structure of Twitter, the baseline is G.
\end{itemize}

Baseline methods may check multiple boxes and have multiple types. For exmample, BotRGCN \citep{feng2021botrgcn} is F since it selects user metadata and encode users as feature vectors. BotRGCN is T since it encodes user tweets and description with pre-trained RoBERTa. BotRGCN is G since it constructs a relational graph and adopts relational graph neural networks for bot detection. As a result, BotRGCN has the type of FTG.

\subsection{F1-score Results}
We re-implement 35 Twitter bot detection baselines and evaluate them on 9 representative datasets and benchmarks. We present their detection accuracy in Table \ref{tab:big_acc} and F1-score in Table \ref{tab:big_f1}.

\begin{table}[t]
    \centering
    \caption{Example hashtags in the five hashtag-based sub-communities.}
    \resizebox{\linewidth}{!}{
    \begin{tabular}{c|c}
        \toprule[1.5pt]
        \textbf{ID}&\textbf{Example Hashtags}\\
        \midrule[1pt]
        \multirow{3}{*}{1}&
        \multirowcell{3}{
        \#Christ, \#Taliban, \#Kabul, \#Germany, \#EU, \#manufacturer, \#Manchester, \#Covid, \#covid, \#bitcoin, \\\#Ukraine, \#Kyiv, \#Iowa, \#farm, \#health, \#bullying, \#Putin, \#gerrymandering, \#Covid19, \#Labour}\\
        &\\
        &\\
        \hline
        \multirow{3}{*}{2}&\multirowcell{3}{\#cybersecurity, \#CVE, \#GCP, \#marketing, \#datacenter, \#OSINT, \#SMEs, \#aerospace, \#innovation, \#science, \\\#exoplanet, \#log4j, \#conservation, \#farming, \#biology, \#chemistry, \#agriculture, \#growth, \#aging, \#dementia}\\
        &\\
        &\\\hline
        \multirow{3}{*}{3}&\multirowcell{3}{\#Curitiba, \#Colombia, \#inversiones, \#emprender, \#emprendedor, \#negocios, \#liderazgo, \#bici, \#ciclismo, \#RRSS, \\ \#correr, \#sueños, \#metas, \#familia, \#inversión, \#Fortalecimiento, \#emprendimiento, \#Familia, \#éxito, \#ventas}\\
        &\\
        &\\\hline
        \multirow{3}{*}{4}&\multirowcell{3}{\#Vimeo, \#Industry, \#Anonymous, \#iHeartRadio, \#Biomass, \#Contest, \#Books, \#Humor, \#Memoir, \#Storytelling,\\ \#Butterfly, \#Art, \#Canvas, \#Handbag, \#Tshirt, \#Kidney, \#Passion, \#Quarantine, \#Whitelist, \#PMC}\\
        &\\
        &\\\hline
        \multirow{3}{*}{5}&\multirowcell{3}{\#UCL, \#coach, \#FF, \#USMNT, \#Coventry, \#Orpheus, \#CRO, \#Sydney, \#Houston, \#Jordan,\\ \#Buffalo, \#UBC, \#writer, \#Shona, \#Christchurch, \#Antigua, \#Sonny, \#Gladstone, \#500th, \#Philips}\\
        &\\
        &\\
        \bottomrule[1.5pt]
    \end{tabular}}
    
    \label{tab:hashtag}
\end{table}

\subsection{Graph Component Removal Details}
\label{app:graph_removal}
We remove the graph component in graph-based methods to examine the role of graphs in Twitter bot detection and present results in Table \ref{tab:remove_graph_all}. We provide details about how graphs are removed from each baseline as follows:
\begin{itemize}[leftmargin=*]
    \item \textbf{Alhosseini \textit{et al.}}. We remove the GCNs while using two MLP layers with user features.
    \item \textbf{Moghaddam \textit{et al.}}. We remove 11 graph-based features from the "friend preference" section.
    \item \textbf{Knauth \textit{et al.}}. We remove 2 graph-based features, namely friend count and follower count.
    \item \textbf{EvolveBot}. We remove 4 graph-based feature extracted with the help of neighbor information.
    \item \textbf{BotRGCN} and \textbf{RGT}. We remove the R-GCN and RGT while using two MLP layers with user features for bot detection. BotRGCN and RGT use the same user features so that their "w/o graph" results are identical.
\end{itemize}

\begin{table}[]
    \centering
    \caption{Statistics of the 10 sub-communities.}
    \resizebox{1\linewidth}{!}{
    \begin{tabular}{c|c|c|c|c|c|c|c|c|c|c}
        \toprule[1.5pt]
        \textbf{Sub-communities}&0&1&2&3&4&5&6&7&8&9\\
        \midrule[1pt]
        \textbf{\# Human}&5,000&5,000&5,000&5,000&5,000&5,000&5,000&5,000&5,000&5,000\\
        \textbf{\# Bot}&5,000&5,000&5,000&5,000&5,000&5,000&5,000&5,000&5,000&5,000\\
        \textbf{\# User}&10,000&10,000&10,000&10,000&10,000&10,000&10,000&10,000&10,000&10,000\\
        \textbf{\# Tweet}&969,979&942,020&1,099,962&989,536&1,083,655&1,156,640&1,333,018&1,138,480&1,151,362&1,142,717\\
        \textbf{\# Edge}&1,116,208&1,120,637&1,245,190&1,167,285&1,249,535&1,535,397&1,924,616&1,508,054&1,511,824&1,526,627\\
        \bottomrule[1.5pt]
    \end{tabular}
    }
    \label{tab:sub-communities}
\end{table}

\subsection{Generalization Study Details}
\label{app:generalization}
To evaluate existing methods and their ability to generalize on unseen data, we identify 10 sub-communities in the TwiBot-22 network and conduct experiments in Figure \ref{fig:heat}. Specifically, we firstly select 5 closely connected sub-communities around \textit{@BarackObama}, \textit{@elonmusk}, \textit{@CNN}, \textit{@NeurIPSConf}, and \textit{@ladygaga}. These five users feature different interest domains and their neighborhood represents different aspects of the Twitter network. In addition, we use K-means to cluster the word2vec \citep{mikolov2013efficient} representations of hashtags and identify users tweeting about similar hashtags into 5 sub-communities. Examples of these hashtags in these sub-communities are presented in Table~\ref{tab:hashtag}. The statistics of the 10 sub-communities are presented in Table~\ref{tab:sub-communities}.

\subsection{Computation Details}
We ran all experiments on a server with 8 GeForce RTX 2080 Ti GPUs. We run each experiment for five times and report the average model performance as well as standard deviation.

\subsection{Annotation Bias Test}
To study the effect of individual labeling function, we remove each of them and examine how many labels have changed in the snorkel-based annotation process, as is shown in Table \ref{tab:anno_study1}. 
\begin{table*}[]
\renewcommand\arraystretch{1.3}
    \caption{We remove labeling functions in the annotation process and compare their results with the full annotation model.}
    \centering
    \resizebox{1\linewidth}{!}{
    \begin{tabular}{c|ccccc}
        \toprule[1.5pt]
        \textbf{labeling function} & \textbf{bot->bot} & \textbf{bot->human} & \textbf{human->human} & \textbf{human->bot} & \textbf{changed percentage}\\
        \midrule[0.75pt]
        w/o adaboost & 77,050 & 49,965 & 869,317 & 3,986 & 5.36\%\\
        w/o random forest & 117,191 & 9,524 & 841,531 & 31754 & 4.13\%\\
        w/o MLP & 109,152 & 17,563 & 796,159 & 77,126 & 9.46\%\\
        w/o GCN & 120,925 & 5,790 & 833,776 & 39,509 & 4.54\%\\
        w/o GAT & 123,397 & 3,318 & 824,436 & 48,849 & 5.22\%\\
        w/o RGCN & 123,676 & 3,042 & 819,293 & 53,970 & 5.70\%\\
        w/o verify & 122,177 & 4,538 & 873,101 & 184 & 0.47\%\\
        w/o keywords & 123,880 & 2,835 & 840,142 & 33,143 & 3.60\%\\
         \bottomrule[1.5pt]
    \end{tabular}
    }

    \label{tab:anno_study1}
\end{table*}

Experiment results on using the expert labels as the test set is presented in Table \ref{tab:anno_study2}.
\begin{table*}[]
\renewcommand\arraystretch{1.3}
    \caption{We use the training set and validation set in TwiBot-22 while using the expert labels as the test set. We used 6 baseline methods for a quick evaluation. Test 1 indicates using only the 1,000 manually annotated users in Section \ref{appendix: test1}, and test 2 indicates using only the 500 manually annotated users in Section \ref{appendix: test2}.}
    \centering
    \resizebox{1\linewidth}{!}{
    \begin{tabular}{ccccccccc}
        \toprule[1.5pt]
        \multirow{2}{*}{\textbf{Model}} & \multicolumn{4}{c}{\textbf{test set 1}} & \multicolumn{4}{c}{\textbf{test set 2}}\\
        \cmidrule(r){2-5} \cmidrule(r){6-9}
        & \textbf{Accuracy} & \textbf{F1-score} & \textbf{Precision} & \textbf{Recall} & \textbf{Accuracy} & \textbf{F1-score} & \textbf{Precision} & \textbf{Recall}\\
        \midrule[0.75pt]
        Moghaddam \textit{et al.} & $89.41~(\pm 0.30)$ & $24.98~(\pm 2.72)$ & $16.57~(\pm 1.97)$ & $50.79~(\pm 4.25)$ & $83.93~(\pm 0.28)$ & $18.49~(\pm 0.95)$ & $11.58~(\pm 0.59)$ & $45.94~(\pm 3.35)$ \\
        SGBot & $91.87~(\pm 0.11)$ & $47.43~(\pm 1.21)$ & $76.16~(\pm 2.31)$ & $34.48~(\pm 1.56)$ & $87.42~(\pm 0.31)$ & $26.00~(\pm 2.80)$ & $54.55~(\pm 2.80)$ & $17.11~(\pm 2.28)$\\
        BotHunter & $91.44~(\pm 0.12)$ & $40.39~(\pm 0.32)$ & $78.28~(\pm 3.11)$ & $27.24~(\pm 0.52)$ & $85.63~(\pm 0.31)$ & $23.38~(\pm 1.55)$ & $73.67~(\pm 9.81)$ & $13.95~(\pm 1.18)$\\
        GAT & $91.14~(\pm 0.45)$ & $47.00~(\pm 2.92)$ & $64.83~(\pm 4.31)$ & $36.95~(\pm 3.04)$ & $84.93~(\pm 0.23)$ & $30.47~(\pm 2.64)$ & $55.64~(\pm 2.02)$ & $21.05~(\pm 2.46)$\\
        BotRGCN& $88.74~(\pm 0.29)$ & $65.89~(\pm 1.62)$ & $79.82~(\pm 2.53)$ & $56.23~(\pm 3.24)$ & $85.59~(\pm 0.68)$ & $55.45~(\pm 2.77)$ & $67.45~(\pm 2.74)$ & $47.17~(\pm 3.65)$  \\
        RGT & $92.80~(\pm 0.45)$ & $23.39~(\pm 4.61)$ & $ 58.33~(\pm 11.78)$ & $16.44~(\pm 2.98)$ & $87.10~(\pm 1.19)$ & $38.02~(\pm 7.21)$ & $58.50~(\pm 10.18)$ & $28.57~(\pm 6.68)$\\
         \bottomrule[1.5pt]
    \end{tabular}}
    \label{tab:anno_study2}
\end{table*}

\subsection{Implementation Details}
The TwiBot-22 evaluation framework is built with help of many valuable scientific artifacts, including pytorch \citep{pytorch}, pytorch lightning \citep{Falcon_PyTorch_Lightning_2019}, pygod \citep{liu2022pygod}, transformers \citep{wolf-etal-2020-transformers}, pytorch geometric \citep{fey2019pytorchgeometric}, sklearn \citep{scikit-learn}, gensim \citep{gensim}, spacy \citep{Honnibal_spaCy_Industrial-strength_Natural_2020}, tweepy \citep{roesslein2009tweepy}, pandas \citep{mckinney2011pandas}, numpy \citep{harris2020array}, vaderSentiment \citep{hutto2014vader}, and imbalanced-learn \citep{imblearn}.

\end{document}